\documentclass[twoside,12pt]{report} 
\usepackage{epsf,graphicx}
\usepackage{latexsym,amssymb}
\usepackage{cite}

\usepackage{latexsym,a4wide} 
\usepackage{amsfonts} 
\usepackage{amsbsy} 
\usepackage{amssymb} 

\usepackage{setspace,cite} 
\usepackage{fancyhdr}

\usepackage{vmargin}
\setpapersize{A4}
\setmargins{32mm}{20mm}{154mm}{240mm}{12pt}{20pt}{12pt}{36pt}

\newcommand{\C}[1]{{\mathcal #1}}

\newcommand{\beq}{\begin{equation}}
\newcommand{\eeq}{\end{equation}}
\newcommand{\bea}{\begin{eqnarray}}
\newcommand{\eea}{\end{eqnarray}}

\newcommand{\Tr}{{\hbox{Tr}\,}}
\newcommand{\comm}[2]{\left[#1,#2\right]}
\newcommand{\italic}[1]{\begin{it}#1\end{it}}
\newcommand{\absval}[1]{{\left\vert#1\right\vert}}

\newcommand{\thetafn}[1]{{\,\theta \, \!#1}}
\newcommand{\deltafn}[1]{{\,\delta \, \!#1}}

\newcommand{\half}{{1\over 2}}

\newcommand{\quarter}{{1\over 4}}

\newcommand{\nn}{\nonumber}

\def\NbyN{N\! \times \! N}
\def\Pf{\C P}
\def\const{ \hbox{\it{const}}}

\def\Tr{{\rm Tr}}
\def\degree{{\rm deg}}
\def\const{{\rm const.}}

\def\Htil{\widetilde{H}}
\def\Qtil{\widetilde{Q}}
\def\Ctil{\widetilde{C}}
\def\phibar{\overline{\phi}}

\def\gammabar{\overline{\gamma}}

\def\psibar{\overline{\psi}}
\def\NbyN{N\!\! \times \!\! N}
\def\half{{1 \over 2}}
\def\quarter{{1 \over 4}}
\def\phibar{\overline{\phi}}
\def\delbar{\overline{\delta}}

\def\Sdef{S_{\rm{def}}}
\def\Zdef{\C Z_{\rm{def}}}
\def\Pfdef{\C P_{\rm{def}}}

\def\NbyN{N\! \times\!N}

\def\real{\rm{Re}}

\begin{document}



\newpage
\thispagestyle{empty}

\thispagestyle{empty}
\begin{center}
{\Large\bf Yang-Mills Matrix Theory\\}
\vspace{1.3cm}
{
Peter Austing\\
\vspace{0.5cm}
\begin{it}
Department of Physics, 
University of Oxford\\
Theoretical Physics,\\
1 Keble Road,\\
Oxford OX1 3NP, UK
\end{it}
}

\vspace{0.5cm}
{p.austing@physics.ox.ac.uk}
\vspace{1.5cm}
\singlespacing
\bf Abstract
\vspace{0.5cm}
\end{center}
\setlength{\parindent}{0mm}
\setlength{\parskip}{0.5\baselineskip}
In this thesis, we discuss bosonic and supersymmetric Yang-Mills
matrix models with compact semi-simple gauge group.

We begin by finding convergence properties for the models. In the
supersymmetric case, we show that the partition function converges
when $D=4,6$ and $10$, and that correlation functions of degree
$k<k_c=2(D-3)$ are convergent independently of the group. In the
bosonic case we show that the partition function is convergent when $D
\geq D_c$, and that correlation functions of degree $k<k_c$ are
convergent, and calculate $D_c$ and $k_c$ for each group.

We move on to consider the supersymmetric theories in both their
Yang-Mills and cohomological formulations. Specialising to the case of
$SU(N)$ with large $N$, we find all quantities which are invariant
under the 
supercharges.

Finally we apply the deformation method of Moore, Nekrasov and
Shatashvili directly to the supersymmetric Yang-Mills model with
$D=4$. We find 
a deformation of the action which generates mass terms for all the
matrix fields whilst preserving some supersymmetry. This allows us to
rigorously integrate over a BRST quartet and arrive at the well known
formula of MNS.

\setlength{\parindent}{6mm}
\setlength{\parskip}{0\baselineskip}

\vspace{3cm}
\begin{center}
Thesis submitted for the Degree of Doctor of Philosophy
in the\\
 University of Oxford\\
\vspace{0.3cm}
June 2001
\end{center}
\chapter*{}
\thispagestyle{empty}
\begin{center}
\large \bf {Acknowledgments}
\end{center}
\vspace{0.5cm}
I am  lucky to have been supervised by Dr John Wheater for this
project, and I thank him for his guidance and insights, and
above all for his patience and kindness. I am
also  grateful recently to Dr Graziano Vernizzi for illuminating
conversations on
matrix theory, and especially to Dr Martin Harris for his insights
and valuable collaboration towards the beginning of the work. I am
indebted to many other friends and colleagues in the department
and in Oxford who have made my time here very enjoyable. I am grateful
to PPARC for financial support.

A part of this work was carried out at the departments of Professor
H. Kawai at Kyoto University, and Professor S. Iso at KEK, and I
warmly thank them and all of their colleagues and students for their 
tremendous hospitality, and Monbusho for making this possible.

To my family Jennifer, Richard and John, and to Hiroe I owe very
special thanks.
\pagebreak

\doublespacing

\pagenumbering{roman}
\setcounter{page}{1} \pagestyle{plain}

\tableofcontents
\pagebreak


\pagestyle{fancy}

\newpage

\pagenumbering{arabic}
\setcounter{page}{1} \pagestyle{fancy}
\renewcommand{\chaptermark}[1]{\markboth{\chaptername%
\ \thechapter:\,\ #1}{}}
\renewcommand{\sectionmark}[1]{\markright{\thesection\,\ #1}}

\addtolength{\headheight}{3pt}
\fancyhead{}
\fancyhead[RE]{\sl\leftmark}
\fancyhead[RO,LE]{\rm\thepage}
\fancyhead[LO]{\sl\rightmark}
\fancyfoot[C,L,E]{}
\pagenumbering{arabic}

\onehalfspacing

\chapter{Introduction} 
\label{chap:intro}
In this chapter, we give a brief introduction to Yang-Mills matrix
models and discuss some of their applications in modern physics.

\section{Yang-Mills Gauge Theories}The Yang-Mills matrix models
are related to  gauge theories by 
dimensional reduction. In this section, we recall the structure of
Yang-Mills gauge theories, and this will allow us to set up the
notations used throughout the thesis\footnote{In this section we follow
the notation and conventions of \cite{GSW}.}. 

A Yang-Mills gauge theory in $D$
dimensions has 
Lagrangian 
\beq\label{lag}
\C L =  \quarter F^2 -{i \over 2} \psibar \, \gamma \cdot \C D
\psi.
\eeq
The fields in this theory are the vector potential $X_\mu^a$, and
fermions $\psi_\alpha^a$.
Here $\mu$ is a spacetime index running from
$0$ up to $D-1$, and for the moment we use a Minkowski metric. 
The $\gamma^\mu_{\alpha \beta}$ are Dirac matrices, and $\psibar$ is
defined
\beq
\psibar = \psi^{\dagger} \gamma^0.
\eeq
The
fields are in the Lie algebra of a compact semi-simple gauge group $G$
so we 
can write 
\beq
X_\mu = X_\mu^a t^a \, \;\;\;\;\; \psi_\alpha = \psi_\alpha^a t^a
\eeq
where the $t^a$ $(a=1, \cdots , g)$ are the generators of the Lie
algebra which we 
choose such that
\beq
\Tr t^a t^b = 2 \delta^{ab}
\eeq
and
\beq
\comm{t^a}{t^b} = i f^{abc} t^c.
\eeq
The gauge field strength $F$ is defined
\beq
F_{\mu \nu}^a = \partial_\mu X_\nu^a - \partial_\nu X_\mu^a + c
f^{abc}X_\mu^b  X_\nu^c
\eeq
and the gauge covariant derivative is
\beq
(\C D_\mu \psi)^a = \partial_\mu \psi^a + cf^{abc}X_\mu^b \psi^c.
\eeq
The parameter $c$ is a coupling constant. The theory is invariant
under gauge transformation
\beq
\psi \rightarrow U \psi U^{-1} \, , \;\;\;\;\; X_\mu \rightarrow U
X_\mu U^{-1} -i c^{-1} (\partial_\mu U) U^{-1}
\eeq
where $U \in G$.
The fermions are optional in this model. We can define a purely
bosonic gauge theory by simply omitting them.

\section{Spinors and Supersymmetry}
Concentrating on those theories that contain fermions, we now recall
the spin structure. For the present, we continue to use 
Minkowski metric $\eta^{\mu \nu} = \hbox{diag}(-1, +1,  \cdots ,
+1)$. In any dimension $D$, we can find an irreducible representation
of the
Dirac matrices $\gamma^\mu_{\alpha \beta}$ which satisfy the Clifford
algebra\footnote{In this section, we follow \cite{QFT}.} 
\beq
\{ \gamma^\mu , \gamma^\nu \} = 2 \, \eta^{\mu \nu} \, 1.
\eeq
Then, defining
\beq
S^{\rho \sigma} = \quarter \comm{\gamma^\rho}{\gamma^\sigma}
\eeq
one can verify that these matrices satisfy the Lorentz algebra
\beq
\comm{S^{\rho \sigma}}{S^{\tau \upsilon}} = \eta^{\sigma \tau} S^{\rho
\upsilon} - \eta^{\rho \tau} S^{\sigma \upsilon} + \eta^{\rho
\upsilon} S^{\sigma \tau} - \eta^{\sigma \upsilon} S^{\rho \tau}.
\eeq
On the other hand, we have the Lorentz generators
\beq
(M^{\sigma \rho})^\mu_\nu = \eta^{\rho \mu} \delta^\sigma_\nu -
\eta^{\sigma \mu} \delta^\rho_\nu
\eeq
which act on the ordinary spacetime indices carried by $X^\mu$. These
two representations of the Lorentz algebra are inequivalent. We define
Dirac spinors $\psi_\alpha$ to transform as
\beq
\psi \rightarrow \psi + \omega_{\rho \sigma} S^{\rho
\sigma}  \psi
\eeq
whilst $X^\mu$ transforms as
\beq
X^\mu \rightarrow X^\mu + \omega_{\rho \sigma} (M^{\rho
\sigma})^\mu_\nu X^\nu
\eeq
under an infinitesimal Lorentz transformation. The nice property
\beq
\comm{\gamma^\mu}{S^{\rho \sigma}} = (M^{\rho \sigma})^\mu_\nu \,
\gamma^\nu
\eeq
shows that the Dirac matrices are invariant under Lorentz
transformations, and so 
the Lagrangian \ref{lag} transforms as a  Lorentz
scalar. 

When $D$ is even, one can define
\beq
\gammabar = (-1)^{\quarter(D+2)} \gamma^0 \gamma^1 \cdots
\gamma^{D-1}
\eeq
which has the properties
\beq
\gammabar^2=1 \, , \;\;\; \{ \gammabar , \gamma^\mu \} =0 \, , \;\;\;
\comm{\gammabar}{S^{\rho \sigma}} =0.
\eeq
Then $\gammabar$ is Lorentz invariant, and has eigen-values $+1$ and
$-1$. In this case, we can define \begin{it}Weyl spinors\end{it}  by
projecting onto one 
of these eigen spaces.

If it is possible to choose the Dirac matrices to be all real or all
imaginary, then one can impose the spinors to be real and still
preserve Lorentz symmetry. This is called the Majorana condition, and
is possible when $D= 0,1,2,3 \hbox{ or } 4 \hbox{ (mod 8)}$. (Of
course, it 
may be
more convenient to work in a different basis in which the Dirac
matrices are not 
all real or all imaginary, and in that case one needs to appropriately
modify the Majorana condition.)
Precisely when $D=2 \hbox{ (mod 8)}$, it is possible to apply both the
Weyl and 
Majorana conditions.

We are particularly interested in the possible dimensions and spinor
types for which the theory \ref{lag} may be supersymmetric. The
crucial constraint is that there must be the same number of physical
bosonic modes (that is $(D-2)g$) as physical fermions. The size of the minimal
representation of the Dirac matrices increases very rapidly with $D$,
so this can only be possible for certain small $D$. In fact it is
possible in $D=3$ with a Majorana spinor, $D=4$ with a Weyl (or
Majorana) spinor,
$D=6$ with a Weyl spinor and $D=10$ with a Majorana-Weyl spinor. 
All of these theories are invariant under the supersymmetry
\beq\label{intro:ss}
\begin{array}{ccc}
\delta X_\mu^a &=& {i \over 2} \overline{\xi} \gamma_\mu \psi^a\\
\delta \psi^a &=& -  F^a_{\mu \nu} S^{\mu \nu} \xi
\end{array}
\eeq
as long as one  applies the equations of motion. This may not seem
like much help, but it is possible to
introduce some auxiliary fields to \ref{lag}, and then write down a
supersymmetry which is also valid off shell. We shall do this
explicitly for $D=4$ in chapter \ref{chap:coho}.

Each
of these theories has $(D-2)g$ physical fermionic degrees of freedom. The
spinor index $\alpha$ runs from $1$ to $2(D-2)$ in the case of a real
(Majorana) representation, and from $1$ to $D-2$ in the case of a
complex representation. Then we
see from \ref{intro:ss} that there are $\C N = 2(D-2)$ real
supercharges.

\section{Yang-Mills Matrix Models}
To obtain a Yang-Mills matrix model, we take the Lagrangian \ref{lag}
and assume all the fields are independent of space and
time. Effectively, this means we drop all the derivative terms from
\ref{lag}. 

At this stage, we also move from Minkowski to Euclidean signature. We
do this by setting
\beq
X_0^a = i X_D^a \, , \;\;\;\; (a=1, \cdots , g)
\eeq
and taking the $X_D^a$ real. We also set
\beq
\gamma^0 = i \gamma^D.
\eeq
We have been careful to leave this manipulation until last because we
wish to study the ``Wick rotation'' of a Minkowski theory\footnote{We
certainly do not attempt to give any justification for performing a
Wick rotation here. However, the reader may take some
solace from the fact 
that, in one case, it is precisely this model that appears
physically. This is 
the Yang-Mills integral as the bulk part of a Witten index, which we
shall discuss  a little 
more later.}. This leads to a rather strange effect in the case of
$D=10$ when the fermions are Majorana. Since the Dirac matrices can no
longer all be imaginary, an $SO(D)$ transformation would break the
Majorana condition. However, after integrating out the fermions, full
$SO(D)$ invariance is restored since we can analytically
continue in $X_D$. 

We arrive at the matrix model action
\beq\label{liberty1}
S_{Y \! M} = -\Tr \left( \quarter
\comm{X_\mu}{X_\nu}\comm{X_\mu}{X_\nu} + \half \psibar \gamma^\mu
\comm{X_\mu}{\psi} \right)
\eeq
where we have dropped the coupling constant $c$ since it can be scaled
out in a trivial manner. In those cases where the fermions were
originally Majorana (before Wick rotation), we may choose the
representation in which the $\psi_\alpha^a$ are real.
In those cases where the fermions are complex, it will sometimes
be convenient to re-write the $(\psi_\alpha^a)$ for each $a$
as a real vector of double the length. We can also absorb the
$\gamma^0$ which appears in the definition of $\psibar$ into the
$\gamma^\mu$. Thus we shall sometimes write the action in the form
\beq\label{liberty2}
S_{Y \! M} = -\Tr \left( \quarter
\comm{X_\mu}{X_\nu}\comm{X_\mu}{X_\nu} + \half \psi_\alpha
\Gamma^\mu_{\alpha \beta}
\comm{X_\mu}{\psi_\beta} \right)
\eeq
where the $\Gamma^\mu$ are some new matrices defined in terms of the
$\gamma^\mu$,  and the $\psi_\alpha^a$
are now always real. In the case where the original fermions  were
complex, the range of the indices $\alpha$ and $\beta$ has been doubled.

We define a partition function
\beq  \C Z_{D,G} = \int 
\prod_{\mu=1}^D dX_\mu 
 \prod_{\alpha=1}^{\C{N}} d \psi_{\alpha}
\exp \left( \quarter \sum_{\mu , \nu}
\Tr  \comm{X_\mu}{X_\nu}^2 + \half \Tr \psi_{\alpha} [ \Gamma_{\alpha
\beta}^{\mu} X_{\mu}, \psi_{\beta}] \right) \label{P1}\label{intro:partn}
\eeq
which we shall also sometimes refer to  as the Yang-Mills
integral.
In principle one can integrate out the fermions to obtain
\beq
\label{1.3}
\C Z_{D,G} =\int 
\prod_{\mu=1}^D dX_\mu \,
\Pf_{D,G} (X_\mu )
\exp \left( \quarter \sum_{\mu ,\nu}
\Tr \comm{X_\mu}{X_\nu}^2 \right)
\eeq
where the Pfaffian $\Pf_{D,G}$ is a homogeneous polynomial of degree
$\half \C N g$. In this representation, the gauge symmetry is
\beq X_\mu\to U^\dagger X_\mu U,\qquad U\in G \eeq
and $SO(D)$ symmetry
\beq X_\mu\to \sum_\nu Q_{\mu\nu}X_\nu,\qquad Q\in SO(D).\eeq
In addition, we shall consider simple correlation functions
\beq\label{correl}\label{intro:correln}
<\C C_k (X_\sigma )> \;\; = \int 
\prod_{\mu=1}^D dX_\mu \, \C C_k (X_\sigma ) \,
\Pf_{D,G} (X_\mu )
\exp \left( \quarter \sum_{\mu ,\nu}
\Tr \comm{X_\mu}{X_\nu}^2 \right)
\eeq
with $\C C_k$ a function of the $X_\sigma$ which grows like a
polynomial of 
degree $k$.

We shall study two cases of particular interest. In the first case, we
suppress 
the fermions and consider the purely bosonic model (so that the number
of supercharges is $\C N =0$
and the Pfaffian is just $1$). Since the bosonic action also gives the
exponent in all models which include fermions, it is crucial for
understanding the behaviour of the flat directions and how they can be
suppressed. Secondly, we study the models with supersymmetric
action. They can only be written down when $D=3,4,6$ and $10$, and
have $\C N= 2(D-2)$ real supercharges. In the particular case of
$D=10$, this is the IKKT model of IIB string theory.

The first question one must ask about these models is whether the
integrals \ref{intro:partn} and \ref{intro:correln} which define the
partition function and correlation functions are well
defined. Certainly, we must require at least that the partition function is
finite for the theory to 
make any sense. The difficulty here is that the potential
$\Tr \comm{X_\mu}{X_\nu}\comm{X_\mu}{X_\nu}$ has flat directions in which
the matrices commute. For example, in the bosonic case, one can move
to infinity along one of these 
directions whilst keeping the integrand constant, and thus it was widely
believed that these integrals may be infinite. However, in the case of
$SU(2)$ it is possible to perform the integrals for the partition
function exactly. This was done originally in the supersymmetric cases
\cite{Smilga:1986jg,Yi:1997eg,Sethi:1997pa,Suyama:1998ig} and it was
found that the partition function does converge at least for
$D=4,6,10$. Subsequently,
eigen-value densities and some correlation functions have been
calculated in \cite{Krauth:1999rc}. It was   
believed that the supersymmetric versions should be more convergent
than the bosonic because the contributions from the fermionic
integrals would be close to zero near the flat directions. However,
the $SU(2)$ bosonic partition function was  calculated in
\cite{Krauth:1998xh}, 
and was found to converge when $D \geq 5$.

The authors of \cite{Krauth:1998xh} were able to use Monte Carlo
methods to calculate the supersymmetric integrals numerically for
$SU(2)$ and $SU(3)$, and the calculations have been extended to
various other  gauge groups, and also to the bosonic
theories \cite{Krauth:1998yu, Krauth:2000bv}. A difficulty with
numerical simulations for the
supersymmetric integrals is in performing the fermionic integrations
to obtain the Pfaffian, and for this reason, the exact model has only
been studied for the smaller gauge groups. However, the bosonic models
have now been studied for $SU(N)$ with $N$ up to $768$
\cite{Hotta:1998en,Ambjorn:2000dj}. Analytic approximation schemes
have also been constructed for the bosonic models in \cite{Oda:2000im}
and recently for the $D=4$ supersymmetric model \cite{Sugino:2001}.

The  conclusions of the numerical methods are that the
supersymmetric partition function converges when $D=4,6,10$ and that
the bosonic partition functions converge at least when $D$ is large
enough \cite{Krauth:2000bv}. Chapter \ref{chap:conv} of this thesis
will be  devoted to an  
analytic study of the convergence properties of these integrals.

\section{The IKKT Model of the IIB Superstring}
\label{sec:ikkt}
The supersymmetric Yang-Mills matrix theory with $D=10$ has been
proposed as a constructive definition of  IIB superstring theory
\cite{Ishibashi:1997xs}. We give a very brief introduction here, but
for a review see \cite{Aoki:1998bq}. 

The idea of the IKKT conjecture is to begin with the Green-Schwarz
action for the superstring in the Schild gauge:
\beq\label{schild}
S_{GS} = \int d^2 \sigma \left[ \sqrt{ \hat{g} } \alpha
\left( \quarter \{x^\mu,x^\nu \}^2 -{i \over 2} \psibar \Gamma^\mu 
\{ x^\mu , \psi \} \right) + \beta \sqrt{\hat{g}} \right].
\eeq
Here $\sigma$ are 2-dimensional world-sheet coordinates, $\hat{g} =
\det (\hat{g}_{ab})$ is the determinant of the world sheet metric, and
$\alpha, \beta$ are parameters (which could be scaled out). The
$x^\mu$ are target space coordinates, and the Poisson bracket is defined
\beq
\{ x , y \} = {1 \over \sqrt{ \hat{g}}} \epsilon^{ab}\partial_a x
\partial_b y.
\eeq
The theory is then regularised essentially following a method of
Goldstone and 
Hoppe (for a review, see  
\cite{Taylor:2001vb}).
A function $y$ on the world sheet is replaced
by an $\NbyN$ traceless hermitian matrix $Y$, with a correspondence
\beq
\int d^2 \sigma \sqrt{\hat{g}} \, y \leftrightarrow  \Tr Y
\eeq
and
\beq
\{ x , y \} \leftrightarrow  -i  \comm{X}{Y}.
\eeq
Performing this regularisation, the action \ref{schild} becomes
\beq
S_{IKKT} = -\alpha \left(  \quarter \Tr
\comm{X_\mu}{X_\nu}\comm{X_\mu}{X_\nu} + \half \Tr \psibar
\Gamma^\mu \comm{X^\mu}{\psi} \right) + \beta N 
\eeq
The string partition function
\beq
\int D[x] D[\psi] \exp (-S_{GS})
\eeq
then becomes the matrix integral
\beq\label{ikkt}
 \int \prod_{\mu=1}^D dX_\mu \prod_{\alpha=1}^{16} d\psi_\alpha
\exp( -S_{IKKT})
\eeq
which after scaling out  $\alpha$  becomes the
Yang-Mills matrix  partition function $ \C
Z_{10,SU(N)}$ of equation 
\ref{intro:partn}, with an additional factor $e^{- \beta N}$. 

In their original
proposal, IKKT interpreted the integral over the world sheet metric
$\int D[\hat{g}]$ as a requirement to sum over $N$:
\beq
\C Z_{IKKT} \sim \sum_{N} \C Z_{10,SU(N)} e^{-\beta N}.
\eeq
However, in general, the matrix regularisation procedure outlined above is
valid in the limit $N \rightarrow \infty$, and
 the partition function is often
taken as the large $N$ limit
\beq
\C Z_{IKKT} \sim \lim_{N\rightarrow \infty} \C Z_{10,SU(N)}
\eeq
which would correspond to a more literal application of the Goldstone
Hoppe regularisation. The large $N$ limit is not yet well understood,
and it is not clear exactly how to interpret the model. Nevertheless, an
argument relating Wilson loops in the  matrix model to
string field theory in light-cone gauge provides additional evidence
for the importance of the IKKT model \cite{Fukuma:1998en}.

In principle, one could use the IKKT model to calculate any quantity
in string theory, given enough computer time. In practice though, 
only small $N$ calculations are accessible numerically because of 
difficulty 
in generating the Pfaffian.
Although it is the $D=10$ theory which is relevant for the IKKT model,
it is also possible to study the models in $D=4$ and $D=6$. For $D=4$
the Pfaffian is real and non-negative (see \cite{Ambjorn:2000bf}), and
this has allowed Monte Carlo studies up to $N=48$
\cite{Ambjorn:2000bf, Burda:2000mn,Ambjorn:2000dj}. For $D=6$ and
$D=10$, the Pfaffian is complex in general and standard lattice
methods for dealing with the fermions do not work. 

Since the exact IKKT model is difficult to study numerically, a low
energy effective theory was derived in
\cite{Aoki:1998vn}, and this has been studied for large $N$ in
\cite{Ambjorn:2000dx} by taking the absolute value of the Pfaffian. An
alternative approach has been applied 
\cite{Nishimura:2000wf,Nishimura:2000ds}, in which configurations
which are saddle points of the phase of the Pfaffian have been
studied. In these calculations the authors find the intriguing result
that the integrals are dominated by regions corresponding to a lower
dimension  than $D=10$ (it was suggested in \cite{Aoki:1998vn} that
the dimension $4$ might arise as the natural dimension of a branched
polymer which describes the model). There is also a random surface
approach to the 
IKKT model which has been studied in \cite{Bialas:2000gf}. 

\section{The Matrix Model of M-theory}
\label{sec:bfss}
Shortly before the IKKT model for superstring theory was proposed, a
related model was conjectured as a constructive definition of M-theory
\cite{Banks:1997vh}.  The model corresponds to the quantum mechanics
that one obtains by dimensionally reducing the $SU(N)$ gauge theory
\ref{lag} 
with $D=10$
to one (time) dimension. The proposal is that by taking $N \rightarrow
\infty$, this model
gives M theory in a light cone coordinate system, and has become known as the 
BFSS conjecture (for a review, see \cite{Taylor:2001vb}). After fixing
the gauge $X^0=0$, one obtains the Hamiltonian \cite{Claudson:1985th}
\beq
H = \Tr \left( \half P^iP^i - \quarter \comm{X^i}{X^j}\comm{X^i}{X^j}
- \half \psi \Gamma^i \comm{X^i}{\psi} \right)
\eeq
where the indices $i,j$ run from 1 to 9, and the $P^i$ are the canonical
momenta for the $X^i$. Physical states must also satisfy the gauge
constraint 
\beq
C^a \left| phys \right\rangle =0 \, , \;\;\; a=1, \cdots , \rm{dim}[ su(N)]
\eeq 
where
\beq
C^a = f^{abc}(X_i^bP_i^c - {i \over 2} \psi^b_\alpha \psi^c_\alpha ).
\eeq
An important issue in this model is the question of whether there
exists a unique normalisable vacuum state. This issue can also be
discussed in the $D=4$ and $D=6$ models, and with general gauge
group. It is known from early work
\cite{deWit:1989ct} on supersymmetric Yang-Mills quantum mechanics
that the  models have continuous spectrum of $H$ arbitrarily close to
zero, but the question of an exact vacuum state has remained
unresolved. An approach that has been used is to consider the Witten
Index\footnote{For an alternative approach, see
\cite{Porrati:1998ej}.} which is defined 
\beq
I_W = \lim_{\beta \rightarrow \infty} \Tr (-1)^F e^{-\beta H} =
n^b_0-n^f_0 
\eeq
where the trace is over all physical states, and $F$ is the fermion
number operator. Taking the limit of large
$\beta$ projects onto the zero energy states so that the index gives
the number of bosonic vacuum states minus number of fermionic vacuum
states. If it is true that there is a unique bosonic vacuum state,
then the Witten index must be 1.

In a theory with supersymmetry, equal positive energy bosons and
fermions come in pairs and so we expect
\beq\label{wi}
I(\beta) = \Tr (-1)^F e^{-\beta H}
\eeq
to be independent of $\beta$. If $H$ has a discrete spectrum, one can
prove this very easily, however the proof fails if $H$ has a
continuous spectrum since one would be trying to cancel infinite
quantities inside the trace. The approach of \cite{Yi:1997eg} and
\cite{Sethi:1997pa} is to rewrite the index as
\beq
I_W = I(0) + I^d
\eeq
where $I^d$ is the deficit term $I^d = I(\infty)-I(0)$, and $I(0)$ is
known as the principle or bulk term. In a
supersymmetric 
theory with discrete spectrum $I^d=0$, but we cannot assume this in
the Yang-Mills model.
In order to calculate $I(0)$, the expression \ref{wi} is rewritten as
a path integral on the interval $[0,\beta]$ by introducing a
projection operator onto the gauge 
invariant states. When the limit $\beta \rightarrow 0$ is taken, the
\italic{path} integral becomes an \begin{it}ordinary\end{it} integral
\beq\label{bulk}
I(0) =  {1 \over \C F_{G}}\C Z_{D,G}
\eeq
where $\C Z_{D,G}$ is precisely the supersymmetric matrix partition
function 
of equation \ref{intro:partn} (at least up to some inverse factors of
$2\pi$). The constant $\C F_{G}$ is a group dependent factor and has been
calculated for $SU(N)$ in \cite{Krauth:1998xh}, and some other groups
in \cite{Krauth:2000bv}. The field $X^0$
that is missing from 
the quantum mechanics has become reinstated by the gauge fixing
procedure. Moreover, although the original quantum mechanics has a
Minkowski metric, the matrix $X^0$ appears in \ref{bulk} with
precisely the Wick rotated signature that we have been discussing. 

The authors \cite{Yi:1997eg,Sethi:1997pa} calculated $I(0)$ for
$SU(2)$ and gave an argument for calculating the deficit
term, and 
reached the conclusion that $I_W=1$ for $D=10$ and $I_W=0$ for
$D=4,6$. Then in \cite{Green:1998tn} the argument for calculating the
deficit term was extended to all $SU(N)$ for $D=10$ suggesting
\beq
I^d_{D=10} = - \sum_{m|N, m>1} {1\over m^2}
\eeq
and the arguments have subsequently be extended to other dimensions
and groups \cite{Kac:1999av}.

Thus, the onus appeared to be on calculating the principle part of the
index which is given by the supersymmetric matrix integral
\ref{intro:partn}. Also in \cite{Green:1998tn}, Green and Gutperle made the
conjecture based on $D$-instanton physics
\beq\label{dis10}
\C Z_{10,SU(N)} =  \C F_{SU(N)} \sum_{m|N, m>0} {1\over m^2}.
\eeq
In \cite{Krauth:1998xh}, the constant $\C F_{SU(N)}$ was calculated
and the 
conjecture extended to $D=4,6$
\beq\label{dnot10}
\C Z_{4,SU(N)} = \C Z_{6,SU(N)} = \C F_{SU(N)} {1 \over N^2}.
\eeq
These conjectures have been confirmed for small values of $N$ by the
Monte Carlo evaluations of \cite{Krauth:1998xh,Krauth:1998yu}. The
values \ref{dis10}, \ref{dnot10} also appeared in a very interesting
calculation 
\cite{Moore:1998et} based on
deforming a cohomological action. We shall discuss this in detail in
chapter \ref{chap:3}.

However, the issue of the Witten index has not yet been fully
resolved.  To begin with, there does not yet exist a proof
of the values of the bulk part of the index. Also for at least one
exceptional 
group, 
inconsistencies between the known bulk contribution and conjectured
deficit contribution have been pointed out in
\cite{Staudacher:2000gx}.

\section{Additional Motivation}
The two models of sections \ref{sec:ikkt} and \ref{sec:bfss} give
motivation enough for studying Yang-Mills matrix theories. In
addition, the $D=10$ supersymmetric $SU(N)$ model can be thought of as
a low energy effective theory for $D$-instantons 
($D_{-1}$-branes) \cite{Green:1998yf}. Perhaps rather than 
 as a physical 
theory, the 
$D$-instanton partition function should be regarded as a quantity
which is likely to appear in many stringy calculations. Indeed it was
such a calculation that allowed Green and Gutperle to predict its
value (see section \ref{sec:bfss}).

\section{Thesis Plan}
In chapter \ref{chap:conv} of this thesis we shall consider the
question of convergence of the partition function \ref{intro:partn}
and correlation functions \ref{intro:correln}. We find convergence
conditions for the bosonic and supersymmetric models with any compact
semi-simple gauge group. This is work originally published in
\cite{Austing:2001pk}, and builds on a previous paper
\cite{Austing:2001bd}. 

In chapter \ref{chap:coho} we shall consider the supersymmetric
theories both in their original formulation, and as cohomological
models. Concentrating on the $SU(N)$ models at large $N$, we
completely classify the quantities which are invariant under the
supersymmetry. This is work originally published in
\cite{Austing:2000rm}. 

In chapter \ref{chap:3} we consider how to apply the deformation
method of \cite{Moore:1998et} directly to the supersymmetric
Yang-Mills matrix models. The aim is to use the supersymmetry to
obtain a rigorous exact calculation of the partition function. We find
a deformation of the action that can 
generate mass 
terms for all the fields and still preserve some supersymmetry. This
allows us to integrate over a BRST quartet rigorously, and confirm the
formula that was obtained in \cite{Moore:1998et}. We show why this
method fails so that an alternative regularisation must be
found. However, a proof that the contour prescription of Moore,
Nekrasov and Shatashvili is the correct
regularisation remains elusive.

\chapter{Convergence}
\label{chap:conv}
In this chapter, we establish the convergence properties of Yang-Mills
matrix models. We consider the 
partition function and simple correlation functions in
theories with compact semi-simple gauge group. In the
supersymmetric case, we show that the partition 
function converges when $D=4,6$ and $10$, and that correlation
functions of degree $k < k_c =2(D-3)$ are convergent independently of
the group. In the bosonic case, we show that the partition function
converges when $D \geq D_c$, and that correlation functions of degree
$k<k_c$ are convergent, and calculate $D_c$ and $k_c$ for each
group. The special case of $SU(N)$ establishes the convergence of the
partition function and a set of correlation functions in the IKKT
model of IIB strings.

\section{Convergent Bosonic Integrals}

We consider first the integral \ref{P1} without fermions so that $\C N
=0$ and there is no Pfaffian. The factor $\quarter$ in the original
action can be scaled out in a 
trivial manner, so we drop it here, giving
\beq  \C Z_{D,G} = \int 
\prod_{\mu=1}^D dX_\mu 
\exp \left( \sum_{\mu , \nu}
\Tr \comm{X_\mu}{X_\nu}^2 \right). \label{bospartn}
\eeq
Since the action is built out of
commutators, there are flat directions in which the magnitude of the
$X_\mu$ can be taken to $\infty$ while keeping the integrand
constant. It is these regions which may lead to a
divergence. Therefore it is useful to define a radial variable $R$
giving the magnitude of the $X_\mu$. Let 
\beq X_\mu=Rx_\mu,\quad \Tr x_\mu x_\mu=1 \label{P4}\eeq
where from now on we use the summation convention for repeated
indices.
Noting that
\beq
\Tr x_\mu x_\mu = \sum_{\mu,i,j} \absval{(x_\mu)_{ij}}^2
\eeq
we see that the $x_\mu$ lie on a compact manifold. To rewrite the
integral over the $X_\mu$ in terms of 
$R$ and $x_\mu$, we insert unity
\beq
1 = \int_0^\infty {dR} \,  2R \deltafn{(R^2 - \Tr X_\mu X_\mu )}
\eeq
and scale out $R^2$ from the $\delta$-function.
Then
\beq \C Z_{D, G}=2 \int_0^\infty dR R^{Dg-1} \C X_{D,
G}(R)
\label{P5}\eeq 
with
\beq \C X_{D, G}(R)=\int \prod_{\nu=1}^D dx_\nu \,
\deltafn{\left(1-\Tr x_\mu x_\mu\right)}
\exp\left(-R^4S\right) \label{P6}\eeq
and
\bea S&=&-\Tr \comm{x_\mu}{x_\nu}\comm{x_\mu}{x_\nu}\nn\\
&=&\sum_{i,j,\mu,\nu}\absval{\comm{x_\mu}{x_\nu}_{i,j}}^2.\label{P6.0}\eea
We note that for any  $R$ the integral $\C X_{D, G}(R)$ is bounded by
a constant. If for large $R$
\bea\absval{\C X_{D,
G}(R)}&<&\frac{\const}{R^\nu},\qquad\hbox{with }\:\nu> Dg,\label{P6.1}\eea
then the partition function $\C Z_{D, G}$ is finite. Our tactic
for proving 
convergence of $\C Z_{D, G}$ is  to find a bound of the form
\ref{P6.1} on $\C X_{D, G}(R)$. A sufficient condition for the
correlation function \ref{correl} to converge is obtained by modifying
\ref{P6.1} to require $ \nu > Dg + k$.

From now on, we are only interested in large $R$, so we shall always
assume $R>1$. Let us split the integration region in \ref{P6} into two
\bea \C R_1:&& S
<(R^{-(2-\eta)})^2\nn\\
\C R_2:&& S
\ge(R^{-(2-\eta)})^2\label{P7}\eea
where $\eta$ is small but positive.  We see immediately that the contribution
to $\C I_{D, G}(R)$ from $\C R_2$ is bounded by $A_1\exp(-R^{2\eta})$ (we will
use the capital letters
$A$, $B$ and $C$ to denote constants throughout this chapter)
 and thus automatically satisfies \ref{P6.1}. Thus we can
confine our efforts to the contribution from $\C R_1$. In this region,
we replace
the exponential function by unity and get the total bound
\beq \absval{\C X_{D, G}(R)}<A_1\exp(-R^{2\eta})+\C I_{D, G}(R) 
 \label{P7.1}\eeq
where 
\beq\C I_{D, G}(R) = \int_{\C R_1} \prod_{\nu=1}^D dx_\nu \,
\deltafn{\left(1-\Tr x_\mu x_\mu\right)}. 
 \label{B8}\eeq 
From now on, we shall work with $\C I_{D,G}(R)$, and seek a bound of the
form 
\beq
\label{cond}
\C I_{D,G}(R) <{ \const \over R^\nu}.
\eeq
Then a sufficient condition for the partition function $\C Z_{D,G}$ to
converge 
is
\beq
\label{pconv}
\nu > Dg
\eeq
and for the correlation function \ref{correl} to converge,
\beq
\label{cconv}
\nu > Dg + k.
\eeq
The condition in \ref{P4} means that at least one of the matrices
$x_\mu$ (say $x_1$) must satisfy
\beq
\label{B9}
\Tr x_1 x_1 \geq D^{-1}.
\eeq
It is convenient to express the Lie algebra using the Cartan-Weyl basis
\beq
\label{B11}
\{ H^i, E^\alpha \}
\eeq
where $i$ runs from $1$ to the rank $l$ and $ \alpha $ denotes
a root.
In this basis
\beq
\label{cartanbasis}
\comm{H^i}{H^j}=0\, , \;\;\;\; \comm{H^i}{E^\alpha}=\alpha^i E^\alpha
\eeq
and
\beq
\label{C20}
\begin{array}{llll}
\comm{E^\alpha}{E^\beta}&=& N_{\alpha
\beta}E^{\alpha+\beta}\;\;\;\;\;\; & \hbox{if } \alpha + \beta \hbox{
is a root}\\
& =&2
\absval{\alpha}^{-2}\, \alpha \cdot H& \hbox{if } \alpha = -\beta \\
&=&0& \hbox{otherwise.}
\end{array}
\eeq
Here $E^{-\alpha} = (E^{\alpha})^{\dagger}$, and the normalisation is
chosen  such that 
\beq
\label{innerprod}
\Tr H^i H^j = \delta^{ij} \, , \;\;\; \Tr E^\alpha E^{\beta} =
2  \absval{\alpha}^{-2}\, \delta^{\alpha+ \beta}.
\eeq
Since the integrand and measure are gauge invariant, we can always use
a group element 
to move $x_1$ into the Cartan subalgebra
\beq
x_1 = x^i H^i
\eeq
and reduce the integral over $x_1$ to an
integral over its Cartan modes  \cite{Krauth:2000bv}
\beq
\label{weyl}
\prod_{a=1}^{g} dx_1^a \rightarrow \const \left( \prod_{i=1}^l dx^i
\right) \prod_{\alpha > 0} ( x \cdot \alpha
)^2.
\eeq
Here
\beq
\label{vandermonde}
\Delta^2_G(x) = \prod_{\alpha > 0} ( x \cdot \alpha
)^2
\eeq
is the Weyl measure\footnote{The statement \ref{weyl} due to Weyl is
of course 
non-trivial. It comes from the fact that any $X$ in the Lie algebra
can be written $X=U 
CU^\dagger$ where $C$ is in the Cartan subalgebra and $U$ is a group
element. The integration measure becomes $dX = dU dC J(C)$ where
the Jacobian $J$ is the Weyl measure. Then since the integrand is
gauge invariant, the $U$ integration just gives a constant - loosely the
volume of the group. In the particular case of $SU(N)$ the result is
well known and the Weyl measure is the square of the Vandermonde. I am
grateful to Dr M. Staudacher for  explaining  these
results.}. We
expand the remaining $x_\nu$
\beq
x_\nu = x_\nu^i H^i + x_\nu^\alpha E^\alpha \, \;\;\;\; \nu=2, \cdots
, D
\eeq
with $x_\nu^{-\alpha} = (x_\nu^\alpha)^*$.

Looking back to equation \ref{P6.0}, we certainly have
\beq
-  \Tr
\comm{x_1}{x_\nu}^2 = \absval{ \Tr
\comm{x_1}{x_\nu}^2} < S 
\eeq
and so, in the region $\C R_1$
\beq
-  \Tr
\comm{x_1}{x_\nu}^2 < R^{-2(2-\eta )}
\eeq
for $\nu=2 , \cdots ,D$.
Writing this in terms of the basis \ref{B11} gives
\beq
\label{B15}
4 \sum_{\alpha >0} {(x \cdot \alpha)^2 \over \absval{ \alpha^2}}
\absval{ x_\nu^\alpha}^2 < 
R^{-2(2-\eta )}. 
\eeq
This is the key result because, 
whenever $(x \cdot \alpha )^2$ is bigger than a constant, it gives us
a bound on $x_\nu^\alpha$ and so allows us to bound the integral.

As yet, we have not specified our choice of ordering giving the
concept of positivity for roots. Since there are a finite number of
roots, there is only a finite number of possible choices. In fact, for
any $x$, there is always a choice such that $x \cdot \alpha \geq 0$
whenever $\alpha$ is a positive root. To see this, temporarily fix
$x$ and change basis in the Cartan subalgebra so that $x = (1,0,
\cdots ,0)$. Now follow the usual construction, and define $\alpha$ to
be positive if it's first non-zero element is positive. Then, in
particular, there is a set of $l$ simple roots $\{ s_i \}$ which are
positive, and such that any positive root can be be written $\alpha =
\alpha_i 
s_i$ with the $\{ \alpha_i \}$ non-negative integers. Finally move
back to the original basis. The property $x \cdot \alpha \geq 0$ when
$\alpha>0$ is preserved.

We can now split the integration region into a finite number of
sub-regions; one for each choice of the positive roots. On each
subregion, we have the condition $x \cdot s_i \geq 0$ for $x$. 

The $\{ s_i \}$ form a basis for the $l$-dimensional vectors. We can
define a number $c$ by
\beq
c = \min_{ \{a^2=1\}}  \, \max_{i} \, \absval{a \cdot
s_i}
\eeq
which must be positive.
Then the
condition \ref{B9} tells us that at least one of the simple roots,
$s_1$ say, satisfies $x \cdot s_1 \geq c D^{-\half}$. In addition, any
positive root $\alpha$ which contains the simple root $s_1$ must
satisfy the same relation so that
\beq
\label{C30}
\absval{x \cdot \alpha} \geq c D^{-\half} \hbox{  whenever }\alpha\hbox{ 
contains }s_1.
\eeq

Let us now split up the Lie algebra $\C G$ as follows.
Define $\C G'$ to be the regularly embedded\footnote{A subalgebra is
``regularly 
embedded'' if it is obtained by knocking out some simple roots from
the original algebra.}  subalgebra of $\C
G$ obtained by omitting the simple root $s_1$. Then $\C G'$ has rank
one less than $\C G$ and so there is one Cartan generator $J$  outside
$\C G'$. We can always choose $J$ to commute with $\C
G'$. To see this, note that $s_2, \ldots, s_l$ span an $l-1$
dimensional subspace of the $l$ dimensional root space, so we can
choose a basis in which they all have 
first component zero. In this basis, choose $J=H^1$. Then
\ref{cartanbasis} shows $[J,E^{\alpha'}]=0$ when $E^{\alpha'} \in \C G'$
so that $[J,\C G' ]=0$.
Let us rename the remaining generators of $\C G$ as $\{ F^\beta \}$
where $\beta$ is any root which contains $s_1$. We can summarise some
of the commutation relations as follows:
\beq
\label{bases}
\begin{array}{lll}
\comm{J}{\C G'}&=&0\\
\comm{F^\beta}{\C G'}& \subset& \{ F^\gamma \}\\
\comm{J}{F^\beta}& \subset& \{ F^\gamma \}\\
\comm{\C G'}{\C G'} &\subset&\C G'
\end{array}
\eeq
The first relation of \ref{bases} is given by the construction of $J$. The
other three
relations
follow immediately from \ref{cartanbasis} and \ref{C20}.

\noindent
Expanding
\beq
\label{xdecomp}
x_\mu = y_\mu + \rho_\mu J + \omega_\mu^\beta F^\beta,
\eeq
with $y_\mu \in \C G'$,
the conditions \ref{B15} and \ref{C30} give us a bound
on the $\omega_\mu$.
\beq
\label{B16}
\absval{\omega_\nu^\beta} < {c^{-1} D^{\half} \absval{ \beta^2} \over 4}
R^{-(2-\eta)} \, , \;\;\;\nu = 2, \cdots ,D
\eeq
There are a number of possible choices for $\C G'$ depending on which simple
root has been removed. The correct choice depends first of all on
which of the $x_\mu$ satisfies \ref{B9} (and so is relabeled $x_1$). And
then, given $x_1$, on which of the simple roots satisfies the
condition in \ref{C30} (and so is relabeled $s_1$). Thus, we have split
the integration region up into a finite number of subregions according to the
correct choice of $\C G'$. We shall use \ref{B16} to bound the
integral \ref{B8} in each of  
these regions.
The region giving the least inverse power of $R$ will
then give a bound on $\C I_{D, G}$. 

Let us expand the action in terms of the variables \ref{xdecomp}.
Using the commutation relations \ref{bases} and inner products
\ref{innerprod} we see that the terms linear in $\omega$ vanish giving
\bea
S_{\C G} (x_\mu) &=& S_{\C G'}(y_\mu) + 2 \Tr [y_\mu , y_\nu] [ F^\beta
, F^\gamma ] \omega_\mu^\beta \omega_\nu^\gamma + \Tr \left(
\omega_\nu^\beta [ y_\mu , F^\beta ] - \omega_\mu^\beta [y_\nu ,
F^\beta ] \right. \nonumber \\&& \left. + ( \rho_\mu \omega_\nu^\beta - \rho_\nu
\omega_\mu^\beta ) 
[J, F^\beta ] +\omega_\mu^\beta \omega_\nu^\gamma [F^\beta , F^\gamma
] \right)^2.
\eea
Here we have added suffices to the actions to emphasise that $S_{\C
G}(x_\mu)$ is the original $G$-invariant action whilst $S_{\C
G'}(y_\mu)$ is the $G'$-invariant action.
Since the $y_\mu$ and $\rho_\mu$ are bounded by a constant, this can
be written
\beq
\label{expaction}
S_{\C G}(x) = S_{\C G'}(y) + \C O( \omega^2).
\eeq
Then the bound \ref{B16} on $\omega$ shows that
(up to a trivial scaling constant)
\beq
\label{B18}
x \in \C R_1(G) \Rightarrow y \in \C R_1(G').
\eeq
We shall now take the expression \ref{B8} for $\C I_{D,G}(R)$, restrict the
integration region to that where the appropriate subalgebra is $\C
G'$, and use the preceding results to form a bound and 
integrate out the variables $\rho$ and $\omega$. First, using
\ref{weyl} to reduce the  $x_1$ integral
to  Cartan modes $x$  gives
\beq\label{ICartan}
\C I_{D,G}(R) = \const \int_{\C R_1(G)}  d x
\Delta_G^2(x) \prod_{\nu=2}^D d x_\nu \deltafn{(1-\Tr x_\mu x_\mu )}.
\eeq
Next note that the Weyl measure (\ref{vandermonde}) for 
$\C G$ can be bounded by that for $\C G'$
\beq\label{B20}
\Delta^2_G(x) < \const \, \Delta^2_{G'}(y).
\eeq
This is because when $\alpha$ does not contain $s_1$, $\alpha \cdot x = \alpha
\cdot y$. Thus $\Delta^2_G(x)$ is equal to $\Delta^2_{G'}(y)$ up to
some additional 
factors which  are bounded by a constant. At this stage, we can also
use the inner product relations \ref{innerprod} to
decompose
\beq
\label{deco}
\Tr x_\mu x_\mu = \Tr y_\mu y_\mu + \rho_\mu \rho_\mu +{2
\over \absval{\beta}^2} \absval{\omega_\mu^\beta}^2.
\eeq
It is convenient to rescale the $\omega$ variables to get rid of the
$2/\absval{\beta}^2$ constants, and to use polar coordinates for both
the $\rho_\mu$ and $\omega_\mu^\beta$ variables so that
\beq
\label{radials}
\Tr x_\mu x_\mu = \Tr y_\mu y_\mu + \rho^2 + \omega^2.
\eeq
In polars, the measures become 
\beq
\prod_{\mu=1}^D d \rho_\mu = d\Omega_\rho d\rho \rho^{D-1}
\eeq
and
\beq\label{polar}
\prod_{\stackrel{\mu=2, \cdots , D}{\beta : E^\beta \notin \C G'}} d
\omega_\mu^\beta = \const \, d \Omega_\omega d \omega \omega ^{(D-1)(g-g'-1)-1}
\eeq
Counting the number of $\omega_\mu^\beta$ to get the exponent in
\ref{polar} is crucial to eventually get the correct bound. There is
an $\omega_\mu^\beta$ for each $\mu = 2, \cdots , D$ (but not $\mu =1$ since
$x_1$ was moved into the Cartan subalgebra), and each of the
$F^\beta$. The $F^\beta$ are the generators of $\C G$ which are
neither $J$, nor in $\C G'$, so there are $g-g'-1$ of them. Then the
total number of variables $\omega_\mu^\beta$ is $(D-1)(g-g'-1)$.

Thus, inserting \ref{radials} into \ref{ICartan}, and using the bounds
\ref{B16},
\ref{B18} and \ref{B20} gives
\bea \label{C43}
\C I_{D,G}(R) &<& B_0 \int_{y_\mu \in \C R_1(G')}  d y
\Delta_G^2(y) \prod_{\nu=2}^D d y_\nu   \int_0^{A_0 R^{-(2-\eta)}} d
\omega \\ & &  \; \;\;\;\;\;
\omega^{(D-1)(g-g'-1)-1}  \int_0^1 d \rho \, \rho^{D-1}  \deltafn{(1- \Tr
y_\mu y_\mu - \rho^2 -\omega^2)}. \nonumber
\eea
where $B_0$ and $A_0$ are constants, and we have integrated out the
angular variables $\Omega_\rho$ and $\Omega_\omega$.

Considering the inner integral first, we can integrate $\rho$ out
immediately to obtain
\beq
\half (1- \Tr y_\mu y_\mu
-\omega^2)^{D-2 \over 2} \thetafn{(1-\Tr  y_\mu y_\mu-\omega^2)}.
\eeq
The original
Yang-Mills integral \ref{P1} only makes sense when $D \geq 2$, and in
this case, we can bound the leading factor by a constant. In addition,
we have the bound
\beq
\thetafn{(1-\Tr  y_\mu y_\mu-\omega^2)} \leq \thetafn{(1-\Tr  y_\mu
y_\mu)}
\eeq
so that \ref{C43} can be bounded by
\beq
 B_0 \int_{y_\mu \in \C R_1(G')}  d y
\Delta_G^2(y) \prod_{\nu=2}^D d y_\nu   \int_0^{A_0 R^{-(2-\eta)}} d
\omega 
\omega^{(D-1)(g-g'-1)-1}  \thetafn{(1-\Tr y_\mu y_\mu)}.
\eeq
Finally integrating out $\omega$ gives
\beq\label{B21}
\C I_{D, G}(R) < B_1 R^{-(2-\eta)(D-1)(g-g' -1)} \C F_{D,G'}(R)
\eeq
where
\beq\label{D52}
\C F_{D,G'}(R) =
\int_{\C
R_1(G')} \prod_{\nu=1}^D dy_\nu \, 
\thetafn{\left(1-\Tr y_\mu y_\mu\right)}.
\eeq
Here, since the integrand is gauge invariant,  we have absorbed the
$G'$ Weyl measure and restored the integral
to $G'$ gauge invariant form.
Using the identity
$
\thetafn{ (1 - \Tr y_\mu y_\mu) }= \int_0^1 dt
\deltafn{(t - \Tr y_\mu y_\mu) },
$
and then rescaling $t =
[u/R]^{2-\eta}$ and $y_\mu = \tilde{y}_\mu [u/R]^{1-\eta /2}$ gives
\beq\label{B22}
\C F_{D, G'}(R) = (2-\eta ) R^{-(1-\eta /2)Dg'}
\int_0^R du  \,u^{(1-\eta /2)Dg'-1} \C I_{D,G'}(u).
\eeq
We shall proceed by induction. Our aim is to show that
\beq\label{indint}
\int_0^\infty dR \, R^{Dg-1} \C I_{D,G}(R) < \const
\eeq
If this is true for $G'$, then the integral in
\ref{B22} is bounded by a constant and so
\beq\label{BF}
\C F_{D, G'}(R) < B_2 R^{-(1-\eta /2)Dg'}
\eeq
so that by \ref{B21}
\beq\label{B24}
\C I_{D,G}(R) < B_3 R^{-(1-\eta /2)[2(D-1)(g-g'-1) + Dg']}
\eeq
when $R>1$ (recall that all of our bounds apply only to $R>1$).
The integral \ref{indint} is certainly convergent in the region $0 \leq R
\leq 1$ since $\C I_{D,G}$ is always finite. For $R>1$, we can
substitute the bound \ref{B24} into \ref{indint} and decide whether
\ref{indint} also converges for $G$. 

Our task then is to find the regularly embedded subalgebras $\C G'$ of
$\C G$
and choose the one which leads to the least inverse power of $R$ in
\ref{B24}. Fortunately, the regularly embedded subalgebras can easily
be found
using the Dynkin diagram. The Dynkin diagram for a Lie algebra has a
node for each simple root. The nodes are connected by 3, 2, 1 or 0
lines respectively as the angle between the corresponding roots is
150, 135, 120 or 90 degrees. In addition to these restrictions on
angles, the simple roots of a compact simple Lie algebra can come in
at most two 
different lengths. The notation in this thesis is that nodes
corresponding to shorter roots are coloured black. Knowledge of the
Dynkin diagram is enough to reconstruct the entire Lie algebra. Thus
we can find the regularly embedded subalgebras $\C G'$  by removing
one node from the Dynkin diagram. For an excellent review of Lie
algebra methods including tables of the Dynkin diagrams and dimensions
that we 
shall use in the following, see  \cite{src:difrancesco}.

Before proceeding to consider each group in turn, we make a final
observation. If the regularly embedded subalgebra $\C G'$ is a direct
sum of two (mutually commuting) subalgebras $\C G' = \C G'_1 \oplus \C
G'_2$ then we have
\beq
\C F_{D,G'}(R) < \C F_{D,G'_1}(R) \, \C F_{D,G'_2}(R)
\eeq
since $ \thetafn{(1-\Tr_{G'} y_\mu y_\mu )} \leq \thetafn{(1-\Tr_{G'_1}
y_\mu y_\mu )} \thetafn{(1-\Tr_{G'_2} y_\mu y_\mu )}$. The result
\ref{B24} is  unaffected, but this will help us to deal with the few
$G'$ which have divergent Yang-Mills integrals so that \ref{BF} and
\ref{B24}  are not
true.

We shall now consider each group in turn. We only consider
dimensions $D \geq 3$ since, as we will see in section
\ref{sec:divgt}, the partition function is always divergent when
$D=2$. The case of $SU(r+1)$ is 
most tricky because $SU(2)$ and $SU(3)$ are rather special having
divergent partition function for some low values of $D$. We
work through the $SU(r+1)$ groups in detail to show the method. For
the other groups, one can easily follow the same method, and so we
give less detail. 
\begin{description}
\item[SU(r+1):]
The Dynkin diagram for $su(r+1)$ is
\beq
\unitlength=1mm
\linethickness{0.4pt}
\begin{picture}(64,6)
\put(2,2){\circle{4}}
\put(17,2){\circle{4}}
\put(32,2){\circle{4}}
\put(47,2){\circle{4}}
\put(62,2){\circle{4}}
\put(4,2){\line(1,0){11}}
\put(19,2){\line(1,0){11}}
\put(37,2){\circle*{0.53}}
\put(39.5,2){\circle*{0.53}}
\put(42,2){\circle*{0.53}}
\put(49,2){\line(1,0){11}}
\end{picture}
\eeq
where there are $r$ nodes. To find
the regularly embedded subalgebras $\C G'$
we remove one of the nodes, and discover
\beq
su(r+1) \rightarrow \C G' = su(m) \oplus su(r+1-m) \, , \;\;\; 1 \leq
m \leq r 
\eeq
where we define $su(1) = 0$.
The dimension of $su(m)$ is $m^2-1$, so that
\beq
\label{vals}
g= (r+1)^2-1 \, , \;\;\; \;\;\;  g' = m^2 +(r+1-m)^2 -2.
\eeq
The Lie algebra $su(2)$ has no regularly embedded subalgebra, so
$g'=0$. The arguments leading to \ref{B24} are all still valid (the
only difference is that in this case there are no variables $y_\mu$
corresponding to $\C G'$ so
we know explicitly that the $\C F_{D,\C G'}$ appearing in \ref{B21} is
just a constant). Then setting $g=3$ and $g'=0$ in \ref{B24} gives
\beq
\label{su2}
\C I_{D,SU(2)} < B_3 R^{-(1-\eta /2)4(D-1)}, \;\;\;R>1.
\eeq
Referring back to \ref{indint}, we see that we need
\beq
3D < (1-\eta /2)4(D-1)
\eeq
which can be re-written
\beq
D>4+2 \eta (D-1)
\eeq
for convergence. Of course, this also corresponds to the original condition
\ref{pconv} for the partition function to converge, so by choosing
$\eta$ sufficiently small, 
$\C Z_{D,SU(2)}$ is finite for $D \geq 5$. 

In the cases $D=3$ and $D=4$, we have failed to show that the desired
induction statement \ref{indint} is true. However, 
we can substitute \ref{su2} back into
\ref{B22} to obtain a bound on $\C F_{D,SU(2)}$ even when $D<5$:
\beq
\label{B29}
\C F_{D,SU(2)} < B_4 R^{(1-\eta /2)3D} R^{(1-\eta /2) \delta_{D,3}}
(\log R)^{\delta_{D,4}}, \;\;\; R>1.
\eeq
In dimensions $3$ and $4$, the result is at variance with
\ref{BF}. However, since $\log R$ tends to $\infty$ more slowly than
any positive power of $R$, modification by a $log R$ factor will not affect
any of our conclusions. (We can modify \ref{cond} to add a $\log R$
factor and still leave the conditions \ref{pconv} and \ref{cconv}
unchanged.) Thus it is only for $D=3$ that we must be
careful to use the modified formula.

The Lie algebra $su(3)$ has $su(2)$ as its only regularly embedded
subalgebra. Then substituting \ref{B29} into \ref{B21} gives the bound
\beq
\label{su22}
\C I_{D,SU(3)} < B_3 R^{-(1-\eta /2)[11D-8]} R^{(1- \eta /2 )
\delta_{D,3}} ( \log R )^{ \delta_{D,4}}
\eeq
and we discover
$\C Z_{D,SU(3)}$ converges for $D \geq 4$. In this case,  the
formula \ref{BF} for $\C F_{D,SU(3)}$ is modified only in the case $D=3$, 
and only by a factor of $log R$ which will not affect our results, and
we may proceed as if the induction statement \ref{indint} were true.

For $SU(r+1)$ with $r \geq 3$, it is a simple exercise to discover
which of the possible $\C G'$ 
gives the least inverse power of $R$ behaviour in \ref{B24}. We
substitute the $g$ and $g'$ of \ref{vals} into \ref{B24}, and choose the value
of $m$ which gives the dominant behaviour. The only
point to remember is that we must include an extra $R^{1-\eta/2}$
factor in the case of $\C G' = su(2) \oplus su(r-1)$ when $D=3$, to
allow for the anomalous behaviour of $\C F_{3,SU(2)}$.

We discover that the dominant behaviour is always obtained when $\C G'
= su(r)$, giving $g'= 
r^2-1$. Then \ref{B24} becomes
\beq
\label{su23}
\C I_{D,SU(r+1)} < B_3 R^{-(1-\eta/2)[D(r^2+4r-1) -4r]}\, , \;\;\; r
\geq 3.
\eeq
Taking $\eta$ small, the condition \ref{pconv}
is met  and so the  partition function
$\C Z_{D,SU(r+1)}$ is convergent for $D \geq 3$ when $r \geq 3$ (and
of course, crucially,  the induction statement \ref{indint} is true).

Finally, comparing the bounds (\ref{su2}, \ref{su22} and \ref{su23}) on
$\C I_{D,G}$ with the condition 
\ref{cconv}, we see that
the correlation function \ref{correl} converges when $k<k_c$ with
\beq
k_c = 2rD -D -4r -\delta_{D,3} \delta_{r,2}, \;\;\; r \geq 1, \, D
\geq 3.
\eeq
In this formula, the cases with $k_c \leq 0$ are those for which the
method fails to prove convergence even of the partition function. In
section \ref{sec:divgt} we shall show that these cases are indeed
divergent. 

\item[SO(2r+1), $\bf{r \geq 2}$:]
The Dynkin diagram for $so(2r+1)$ is
\beq
\unitlength=1mm
\linethickness{0.4pt}
\begin{picture}(64,6)
\put(2,2){\circle*{4}}
\put(17,2){\circle{4}}
\put(32,2){\circle{4}}
\put(47,2){\circle{4}}
\put(62,2){\circle{4}}
\put(3.732,3){\line(1,0){11.536}}
\put(3.732,1){\line(1,0){11.536}}
\put(19,2){\line(1,0){11}}
\put(37,2){\circle*{0.53}}
\put(39.5,2){\circle*{0.53}}
\put(42,2){\circle*{0.53}}
\put(49,2){\line(1,0){11}}
\end{picture}
\eeq
where there are $r$ nodes, and the dimension is $g=2r^2+r$. By
removing one  node, we see 
that the possible $\C G'$ are $so(2m+1) \oplus su(r-m)$ with $0 \leq m
\leq r-1$. We discover the most important contribution is always from
$\C G' = so(2r-1)$, and that $\C Z_{D,SO(2r+1)}$ always converges
for $r \geq 2$ and $D \geq 3$. The critical degree $k_c$ for
correlation functions is
\beq
\begin{array}{lllll}
k_c &=& 2&\;\;\;\;&r=2 \, , D=3 \\
k_c &=& 4rD-8r-3D+4&\;\;\;\;&\hbox{otherwise.}
\end{array}
\eeq
The exception when $r=2$ and $D=3$ occurs because of the anomalous
behaviour of $\C F_{3,SU(2)}$. 

\item[Sp(2r), $\bf{r \geq 2}$:]
The Dynkin diagram for $sp(2r)$ is
\beq
\unitlength=1mm
\linethickness{0.4pt}
\begin{picture}(64,6)
\put(2,2){\circle{4}}
\put(17,2){\circle*{4}}
\put(32,2){\circle*{4}}
\put(47,2){\circle*{4}}
\put(62,2){\circle*{4}}
\put(3.732,3){\line(1,0){11.536}}
\put(3.732,1){\line(1,0){11.536}}
\put(19,2){\line(1,0){11}}
\put(37,2){\circle*{0.53}}
\put(39.5,2){\circle*{0.53}}
\put(42,2){\circle*{0.53}}
\put(49,2){\line(1,0){11}}
\end{picture}
\eeq
where there are $r$ nodes, and the dimension is $g=2r^2+r$. The
possible $\C G'$ are $sp(2m) \oplus su(r-m)$ with $0 \leq m
\leq r-1$, and the dominant contribution is from $sp(2r-2)$. The
partition function $\C Z_{D,Sp(2r)}$ converges for all $r \geq 2$ and
$D \geq 3$ and the critical correlation function is given by
\beq 
\begin{array}{lllll}
k_c &=& 2&\;\;\;\;&r=2 \, , D=3 \\
k_c &=& 4rD-8r-3D+4&\;\;\;\;&\hbox{otherwise.}
\end{array}
\eeq

\item[SO(2r), $\bf{r \geq 4}$:]
The Dynkin diagram for $so(2r)$ is
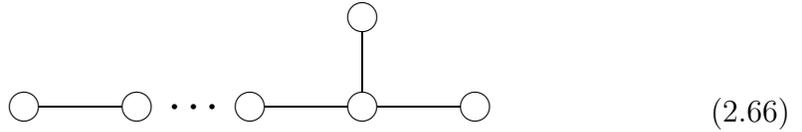
\beq
\unitlength=1mm
\linethickness{0.4pt}
\begin{picture}(64,18)
\put(2,2){\circle{4}}
\put(17,2){\circle{4}}
\put(32,2){\circle{4}}
\put(47,2){\circle{4}}
\put(62,2){\circle{4}}
\put(47,14){\circle{4}}
\put(4,2){\line(1,0){11}}
\put(22,2){\circle*{0.53}}
\put(24.5,2){\circle*{0.53}}
\put(27,2){\circle*{0.53}}
\put(34,2){\line(1,0){11}}
\put(49,2){\line(1,0){11}}
\put(47,4){\line(0,1){8}}
\end{picture}
\eeq
where there are $r$ nodes, and the dimension is $g=2r^2-r$. The
possible $\C G'$ are $so(2m) \oplus su(r-m)$ for $4 \leq m \leq r-1$,
$su(4) \oplus su(r-3)$, $su(r-2) \oplus su(2) \oplus su(2)$ and
$su(r)$. The dominant contribution always comes from $so(2r-2)$, and we
discover that $\C Z_{D,SO(2r)}$ always converges for $D \geq 3$ and $r
\geq 4$. The critical correlation function is given by
\beq
k_c = 4rD -5D -8r +8.
\eeq

\item[$\bf{G_2}$:]
The Dynkin diagram is
\beq
\unitlength=1mm
\linethickness{0.4pt}
\begin{picture}(64,6)
\put(2,2){\circle{4}}
\put(17,2){\circle*{4}}
\put(4,2){\line(1,0){11}}
\put(3.323,0.5){\line(1,0){12.354}}
\put(3.323,3.5){\line(1,0){12.354}}
\end{picture}
\eeq
and the dimension is $14$. The only regularly embedded subalgebra is
$su(2)$, and we discover $\C Z_{D,G_2}$ converges for $D \geq 3$ with
\beq
k_c = 9D - 20 - \delta_{D,3}.
\eeq

\item[$\bf{F_4}$:]
The Dynkin diagram is
\beq
\unitlength=1mm
\linethickness{0.4pt}
\begin{picture}(64,6)
\put(2,2){\circle*{4}}
\put(17,2){\circle*{4}}
\put(32,2){\circle{4}}
\put(47,2){\circle{4}}
\put(4,2){\line(1,0){11}}
\put(18.732,3){\line(1,0){11.536}}
\put(18.732,1){\line(1,0){11.536}}
\put(34,2){\line(1,0){11}}
\end{picture}
\eeq
and the dimension $g=52$. The dominant contributions come equally from
$\C G' = so(7)$ and $\C G' = sp(6)$, each having $g'=21$. Then $\C
Z_{D,F_4}$ converges for $D 
\geq 3$ and
\beq
k_c = 29D-60.
\eeq

\item[$\bf{E_6}$:]
The Dynkin diagram is
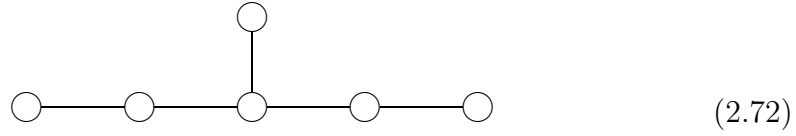
\beq
\unitlength=1mm
\linethickness{0.4pt}
\begin{picture}(64,18)
\put(2,2){\circle{4}}
\put(17,2){\circle{4}}
\put(32,2){\circle{4}}
\put(47,2){\circle{4}}
\put(62,2){\circle{4}}
\put(32,14){\circle{4}}
\put(4,2){\line(1,0){11}}
\put(19,2){\line(1,0){11}}
\put(34,2){\line(1,0){11}}
\put(49,2){\line(1,0){11}}
\put(32,4){\line(0,1){8}}
\end{picture}
\eeq
and the dimension $g=78$. The dominant contribution comes from $\C G'
= so(10)$ having $g'=45$. Then $\C Z_{D,E_6}$ converges for $D \geq 3$
and
\beq
k_c = 31D-64.
\eeq

\item[$\bf{E_7}$:]
The Dynkin diagram is
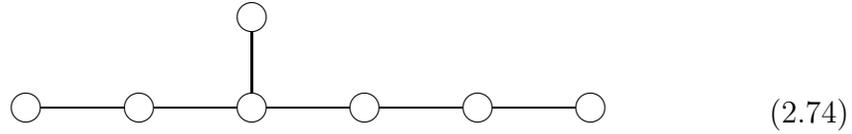
\beq
\unitlength=1mm
\linethickness{0.4pt}
\begin{picture}(79,18)
\put(2,2){\circle{4}}
\put(17,2){\circle{4}}
\put(32,2){\circle{4}}
\put(47,2){\circle{4}}
\put(62,2){\circle{4}}
\put(77,2){\circle{4}}
\put(32,14){\circle{4}}
\put(4,2){\line(1,0){11}}
\put(19,2){\line(1,0){11}}
\put(34,2){\line(1,0){11}}
\put(49,2){\line(1,0){11}}
\put(64,2){\line(1,0){11}}
\put(32,4){\line(0,1){8}}
\end{picture}
\eeq
and the dimension $g=133$. The dominant contribution comes from $\C G'
=e_6$ with $g'=78$. Then $\C Z_{D,E_7}$ converges for $D \geq 3$ and
\beq
k_c = 53D-108.
\eeq

\item[$\bf{E_8}$:]
The Dynkin diagram is
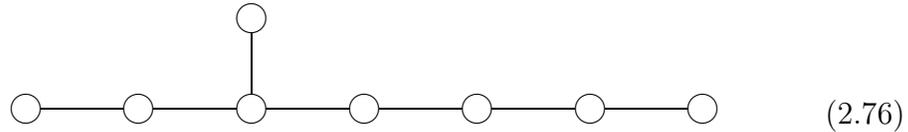
\beq
\unitlength=1mm
\linethickness{0.4pt}
\begin{picture}(94,18)
\put(2,2){\circle{4}}
\put(17,2){\circle{4}}
\put(32,2){\circle{4}}
\put(47,2){\circle{4}}
\put(62,2){\circle{4}}
\put(77,2){\circle{4}}
\put(92,2){\circle{4}}
\put(32,14){\circle{4}}
\put(4,2){\line(1,0){11}}
\put(19,2){\line(1,0){11}}
\put(34,2){\line(1,0){11}}
\put(49,2){\line(1,0){11}}
\put(64,2){\line(1,0){11}}
\put(79,2){\line(1,0){11}}
\put(32,4){\line(0,1){8}}
\end{picture}
\eeq
with dimension $g=248$. The dominant contribution comes from $\C G' =
e_7$ with $g'=133$. Then $\C Z_{D,E_8}$ converges for $D \geq 3$ and
\beq
k_c = 113D-228.
\eeq
\end{description}
\label{sec:conv}

\section{Divergent Bosonic Integrals} \label{sec:divgt}
The lowest $D$ partition function that we can sensibly write
down is for $D=2$,
\beq
\C Z_{2,G} = \int dX_1 dX_2 \exp \left( \Tr \comm{X_1}{X_2}^2 \right).
\eeq
We
can  use the gauge symmetry (by invoking \ref{weyl}) to reduce the $X_1$
integral to Cartan modes, but 
then the integrand is independent of the Cartan modes of $X_2$. Thus,
it is immediate that this integral diverges for every group. From now
on in this section, we assume $D \geq 3$. 

In the previous section \ref{sec:conv} we found an upper bound on $\C
I_{D,G}(R)$. This
equivalently gave us an upper bound on $\C X_{D,G}(R)$
(originally defined in equations \ref{P5} and \ref{P6}). We used the
large $R$ 
behaviour to show that many of  
the partition and correlation functions are finite. 

We shall now find
a lower bound on $\C X_{D,G}(R)$. We shall discover that the large $R$
behaviour of this lower bound is almost identical to that of the upper
bound. The only difference is that the (arbitrarily small) parameter
$\eta$ of the 
previous section is set to zero.

Since the
integrand is positive, it is
sufficient to consider a sub-region of the phase space in order to find a
lower bound. This time, we consider the region
\bea \C R:&& S
<R^{-4}\eea
Then $\exp (-R^4 S) > \exp(-1)$ and so
\beq
\label{D2}
\C X_{D, G}(R) > C_1 \, \C I_{D,G}
\eeq
where now
\beq
\C I_{D,G}=\int_{\C R} \prod_{\nu=1}^D dx_\nu \,
\deltafn{\left(1-\Tr x_\mu x_\mu\right)}
\eeq
and, moving $x_1$ into the Cartan subalgebra,
\beq
\label{D3}
\C I_{D, G}(R) = C_2 \int_{\C R} \prod_{i=1}^l dx_1^i
\Delta^2_G(x_1^i)  \prod_{\nu=2}^D dx_\nu \, 
\deltafn{\left(1-\Tr x_\mu x_\mu\right)}.
\eeq
Now pick a regularly embedded sub-algebra $\C G'$ of $\C G$ (with rank
$1$ less than $\C G$).
As before, write $x =y+
\rho J  + \omega^\beta F^\beta$ with $y \in \C G'$. We will again
write $s_1$ for the simple root of $\C G$ which is removed in order to
obtain $\C G'$. As usual, we use a basis in
which $s_1$ is the only simple root which has its first element
non-zero (indeed, we have already chosen this basis, since we have set
$H^1 =J$ with $\comm{J}{\C G'} =0$).

\noindent
Define a new region $\C R'_\epsilon$ by
\beq
\begin{array}{lll} \label{reg}
\C R'_\epsilon: \;\;\;& \absval{\absval{\omega}} < \epsilon
R^{-2}&  \\
& S_{G'}(y_\mu)  <
\epsilon R^{-4}. 
\end{array} 
\eeq
Then by
taking $\epsilon$ small enough, we see from \ref{expaction}
\beq
R'_\epsilon \subset \C R
\eeq
so that
\beq
\label{D4}
\C I_{D, G}(R) > C_2 \int_{\C R'_\epsilon} \prod_{i=1}^l dx_1^i
\Delta^2_G(x_1^i)  \prod_{\nu=2}^D dx_\nu \, 
\deltafn{\left(1-\Tr x_\mu x_\mu\right)}.
\eeq
As in the previous section, it is convenient to rescale the
$\omega_\nu^\beta$ and write them in polar form (as per equation
\ref{deco}). However, we leave the $\rho_\mu$ as they are for the
moment. Then the integral becomes
\bea\label{clarity}
\C I_{D, G}(R) &>& C_3 \int_{\C R'_\epsilon} \prod_{i=1}^{l-1}dy_1^i
\prod_{\nu=2}^D dy_\nu \prod_{\mu=1}^D d
\rho_\mu \Delta^2_{G}(y_1^i, \rho_1)  \nonumber \\ && d \omega
\omega^{(D-1)(g-g'-1)-1} 
\deltafn{(1- 
\Tr y_\mu y_\mu -\omega^2 -\rho_\mu \rho_\mu )}
\eea
where now the $i$ index runs from $1$ to $l-1$, and $\nu$ runs from
$2$ to $D$.

Choosing to integrate over just two of the $\rho_\mu$, say
$\rho_{D-1}$ and $\rho_D$, leads to
\beq
\int d \rho_{D-1} d \rho_D \deltafn{(1- 
\Tr y_\mu y_\mu -\omega^2 -\rho_\mu \rho_\mu )} 
= C_4 \thetafn{ (
1-\Tr y_\mu y_\mu -\omega^2 - \sum_{\mu=1}^{D-2} \rho_\mu \rho_\mu ) } 
\eeq
as can quickly be seen by writing $\rho_{D-1}$ and $\rho_D$ in
2-dimensional polars. Now, when $R>1$, certainly $\omega^2 <
\epsilon^2$ by \ref{reg}. In addition, we can restrict the region of
integration of each of the remaining $\rho_\mu$ to $-\epsilon <
\rho_\mu < \epsilon$ since we are looking for a lower bound on the
integral. Then we have the inequality
\beq
\thetafn{ (
1-\Tr y_\mu y_\mu -\omega^2 - \sum_{\mu=1}^{D-2} \rho_\mu \rho_\mu ) }
\geq \thetafn{ (
1-\Tr y_\mu y_\mu -\epsilon^2 - (D-2) \epsilon^2) }.
\eeq
The integrand is now independent of $\rho_2 ,\cdots ,\rho_{D-2}$, so
we can immediately integrate them out to
obtain the constant $(2\epsilon)^{D-3}$. The $\theta$-function is also now
independent of $\omega$, so we have the factor
\beq
\int_0^{\epsilon R^{-2}} d \omega \omega^{(D-1)(g-g'-1)-1} =
C_5 R^{-2(D-1)(g-g'-1)}
\eeq
Finally, we can scale each of the
$y_\mu$ and also $\rho_1$ by a
factor $ \sqrt{1- (D-1) \epsilon^2}$ to obtain
\beq \label{D96}
\C I_{D, G}(R) > C_6 R^{-2(D-1)(g-g'-1)}  \int_{\C R(G')}
\prod_{i=1}^{l-1} dy_1^i \prod_{\nu=2}^D dy_\nu
\int_{-\epsilon}^\epsilon d \rho_1 \Delta^2_{G}(y_1^i, \rho_1)  
\thetafn{(1- 
\Tr y_\mu y_\mu)}.
\eeq
Here, the integration region $\C R'_\epsilon$  now applies only to the
$\C G'$ 
variables $y_\mu$. Since $R$  
can be trivially scaled by a constant, we have also dropped the
subscript $\epsilon$ so that the region has become simply $\C R( G')$ in
\ref{D96}. 

Lets use the definition \ref{vandermonde} to decompose the Weyl
measure for $G$ into the part for $G'$ and  
additional factors:
\beq\label{decop}
\Delta^2_{G}(y_1^i, \rho_1) = \Delta^2_{G'}(y_1^i) \prod_{\alpha>0 , \,
s_1 \in \alpha} [\alpha \cdot (\rho_1 , y_1^i)]^2,
\eeq
where $(\rho_1 , y_1^i)$ represents the vector in $l$-dimensional root
space, and the product is over all those positive roots which contain the
simple root $s_1$. Since  $s_1$ is the
only simple root to have its first element non-zero, every factor in
the product of \ref{decop} contains a 
$\rho_1$. Thus, the integral that we are left with over $\rho_1$ is
\beq
\int_{-\epsilon}^\epsilon d \rho_1 \prod_{\alpha>0 , \,
s_1 \in \alpha} [\alpha \cdot (\rho_1 , y_1^i)]^2
\eeq
and can be re-written in a rather more transparent way as
\beq\label{D99}
C_7 \int_{-\epsilon}^\epsilon d \rho_1 \prod_{\alpha>0 , \,
s_1 \in \alpha} (\rho_1 + z_\alpha)^2
\eeq
where the $z_\alpha$ are linear combinations  $z_\alpha =
\alpha^i  y^i_1 / \alpha_0$.

No matter what values the $z_\alpha$ take, the integral \ref{D99} is
always positive. The $y_1^i$ and therefore the $z_\alpha$ lie within a
compact set and so we have the bound
\beq
\int_{-\epsilon}^\epsilon d \rho_1 \prod_{\alpha>0 , \,
s_1 \in \alpha} (\rho_1 + z_\alpha)^2 > \min_{ \{ z_\alpha \} }
\int_{-\epsilon}^\epsilon d \rho_1 \prod_{\alpha>0 , \, 
s_1 \in \alpha} (\rho_1 + z_\alpha)^2 = C_8 >0.
\eeq
Substituting back into \ref{D96} gives
\beq\label{D101}
\C I_{D, G}(R) > C_9 R^{-2(D-1)(g-g'-1)} \int_{\C R(G')}
 \prod_{\nu=1}^D dy_\nu  
\thetafn{(1- 
\Tr y_\mu y_\mu)}
\eeq
where since the integrand is now $G'$ gauge invariant we have absorbed
the $G'$ Weyl measure and restored the integral to full $G'$-gauge invariant
form.

We now follow the method of the previous section (equations \ref{D52},
\ref{B22}) and set
\beq
\C F_{D,G'}(R) =
\int_{\C R(G')}
\prod_{\nu=1}^D dy_\nu \,
\thetafn{\left(1-\Tr y_\mu y_\mu\right)}. 
\eeq
Using the identity
$
\thetafn{ (1 - \Tr y_\mu y_\mu) }= \int_0^1 dt
\deltafn{(t - \Tr y_\mu y_\mu) },
$
and then rescaling $t =
[u/R]^2$ and $y_\mu = \tilde{y}_\mu [u/R]$ gives
\beq\label{D103}
\C F_{D, G'}(R) = 2 R^{-Dg'}
\int_0^R du  \,u^{Dg'-1} \C I_{D,G'}(u).
\eeq
Then \ref{D101} becomes
\beq\label{D104}
\C I_{D, G}(R) > C_{10} R^{-2(D-1)(g-g'-1)-Dg'} \int_0^R du  \,u^{Dg'-1}
\C I_{D,G'}(u). 
\eeq
Comparing with \ref{B21} and \ref{B22} of the previous section, we see
that we have proved essentially a converse result. In those majority
of cases for
which the partition function for $G'$ is finite, it is sufficient to
use the 
bound 
\beq
\int_0^R du  \,u^{Dg'-1}
\C I_{D,G'}(u) > \const \, , \;\;\; R>1
\eeq
giving
\beq\label{D106}
\C I_{D, G}(R) > C_{11} R^{-2(D-1)(g-g'-1)-Dg'}. 
\eeq
In those cases for which the partition function for $G'$ is not
finite, we can find a better bound for $G$ by inductively substituting
the bound found for $G'$ into \ref{D104} and performing the
integral. For example, as we discovered in the previous section, the crucial
exceptional case is when $G'=SU(2)$. Since $su(2)$ has no regularly
embedded subalgebra, the bound \ref{D106} holds as it is, with $g=3$ and
$g'=0$
\beq
\C I_{D,SU(2)} > C_{11} R^{-4(D-1)} \, , \;\;\; R>1.
\eeq
Then
\beq
\begin{array}{lcll}
\int_0^R du  \, u^{3D-1} \C I_{D,SU(2)}(u) &>& C_{12} & D\geq 5\\
&>& C_{13} \log R & D=4\\
&>& C_{14} R & D=3
\end{array}
\eeq
so that, using \ref{D104}, we obtain a better bound for $SU(3)$
\beq
\C I_{D,SU(3)} > C_{15} R^{-8(D-1) -3D} (\log R )^{\delta_{D,4}}
R^{\delta_{D,3}}. 
\eeq

For those simple groups other than $SU(3)$, the bound \ref{D106} is
enough for our purpose.
The final step is that we have to choose the regularly embedded
subalgebra $\C G'$ of $\C G$ which gives the tightest lower
bound. However, we have already performed this task in the previous
section since we chose $\C G'$ to give the least inverse power of
$R$ behaviour.

Lets summarise. In the previous section, we found upper bounds on $\C
X_{D, G}(R)$ which allowed us to deduce that certain partition
functions and correlation functions are finite. These upper bounds
depend on a parameter $\eta $ which can be taken arbitrarily small. In
this section, we have found lower bounds on $\C
X_{D, G}(R)$. The large $R$ behaviour for these lower bounds is
precisely the limit when $\eta \rightarrow 0$ of that for 
the upper bounds.

Thus, using $\C X_{D,G} > C_2 \, \C I_{D,G}$ (\ref{D2}), we can
substitute the lower bounds back into the definition 
\ref{P5} and discover that indeed the partition function is divergent
for $SU(2)$ when $D=3,4$, and for $SU(3)$ when $D=3$. Further, lets
consider the correlation function
\beq
\left\langle \, (\Tr X_\mu X_\mu )^{k/2} \, \right\rangle = \int_0^\infty dR R^{Dg+k-1}
\C X_{D,G}(R).
\eeq
Then this integral diverges when $k \geq k_c$ with the values of $k_c$
quoted in the previous section. So, $k_c$ is indeed the critical value for
correlation functions. Every correlation function with $k<k_c $
converges, and there is always a correlation function with $k=k_c$
which diverges.

\section{Convergent Supersymmetric Integrals}
\label{superconv}
We now move on to consider the supersymmetric integrals (\ref{P1})
which we recall can be written down in dimensions $D=3,4,6$ and $10$
with $\C N =2(D-2)$.
Proceed as for the bosonic integrals to set
\beq
\label{S111}
 \C Z_{D, G}=\int_0^\infty dR  R^{Dg-1} R^{(D-2)g} \C X_{D,
G}(R)
\eeq 
where now
\beq \C X_{D, G}(R)=\int \prod_{\nu=1}^D dx_\nu \, \C P_{D,G}(x_\sigma)
\deltafn{\left(1-\Tr x_\mu x_\mu\right)}
\exp\left(-R^4S\right). \eeq
As before, it is sufficient to consider the region
\beq
\C R_1( \C G):\;\;\; S< R^{-2(2-\eta)}
\eeq
We shall again argue by induction, and
for the induction step to work, we will need to prove the result for
the generalised Pfaffian
\beq
\label{modpfaff}
[\Pf^r_{D,G}(x,R)]^{a_1,\cdots ,a_{2r}}_{\alpha_1,\cdots ,\alpha_{2r}}
= R^{-(2-\eta )2r} \int d \psi \exp (\Tr
\Gamma^{\mu}_{\alpha \beta} \psi_\alpha [x_\mu ,
\psi_\beta ] ) \psi^{a_1}_{\alpha_1} \cdots \psi^{a_{2r}}_{\alpha_{2r}}.
\eeq 
The Pfaffian is modified from the
usual definition by the inclusion of $2r$ fermionic insertions, and a
factor of $R^{-(2-\eta )}$ has been included for each
insertion. The modified Pfaffian can be written down for any $r=0,
\ldots , (D-2)g$. If we set $r=0$ then the
original Pfaffian $\Pf_{D,G}$ is recovered (and is of course
independent of $R$).

The structure of the $\Gamma$ matrices will be irrelevant from now
on (although we shall of course use the fact that their elements are
bounded by a constant). For a more compact notation, we shall suppress the
dependence on $\Gamma$, and on the spinor and group indices,
and write 
\beq
\label{simplepfaff}
\Pf^r_{D,G}(x,R)
= R^{-(2-\eta )2r} \int d \psi \exp (\Tr
\psi [x ,
\psi ] ) \psi^1 \cdots \psi^{2r}.
\eeq 
Then defining
\beq
\C I^r_{D, G}(R) = \int_{\C R_1(\C G)} \prod_{\nu=1}^D dx_\nu \,
\absval{\Pf^r_{D,G} 
(x,R)} \deltafn{\left(1-\Tr x_\mu x_\mu\right)}.
\label{PIdef}
\eeq 
we have
\beq 
\absval{\C X_{D, G}^r(R)}<A_1\exp(-R^{2\eta})+\C I_{D, G}^r(R)
\eeq 
Proceeding as in the bosonic case, we choose the relevant regularly
embedded subalgebra $\C G'$, expand $x_\mu = y_\mu + \rho_\mu J +
\omega_\mu^\beta F^\beta$, and note
\beq \label{omegbound}
\absval{\omega_\nu^\beta} < c^{-1} D^{\half}
R^{-(2-\eta)} \, , \;\;\;\nu = 2, \cdots ,D.
\eeq
Further, write
\beq
\psi =\phi + \xi +  \chi
\eeq
with $\phi \in \C G'$, $\xi = \xi J$ and $\chi = \chi^\beta F^\beta$.
Using the relations \ref{bases}, we find
\beq\label{F12}
\begin{array}{lll}
\Tr \psi \comm{x}{\psi}& =& \Tr \phi \comm{y}{\phi} \\
&&+ \Tr \phi
\comm{\omega}{\chi}+
\Tr \chi \comm{\omega}{\phi} \\
&&
+ \Tr \chi \comm{\omega}{\xi}
+ \Tr \xi \comm{\omega}{\chi}\\
&&+ \Tr \chi
\comm{x}{\chi}
\end{array}
\eeq
where $\rho = \rho J$ and $\omega= \omega^\beta F^\beta$.
Inserting this expression into the definition \ref{modpfaff}, and
expanding part of the exponential gives
\bea
\C P^r_{D,G}(x,R) &=& \int d \phi d \chi d \xi \left( { \xi^1 \cdots
\xi^k \over R^{k(2-\eta)}}{ \phi^1 \cdots \phi^m \over R^{m(2-\eta)}}
{\chi^1 \cdots \chi^n \over R^{n(2-\eta)}} \right) \nonumber \\
&& \times \exp ( \Tr \phi \comm{y}{\phi} + \Tr \phi
\comm{\omega}{\chi} + \Tr \chi \comm{\omega}{\phi} + \Tr \chi
\comm{x}{\chi} ) \nonumber \\
&& \times {1 \over (2(D-2) -k)!} ( \Tr \chi \comm{\omega}{\xi} + \Tr
\xi \comm{\omega}{\chi} )^{2(D-2)-k},
\eea
where $k+m+n=2r$. We first perform the integrals over the $\C
N=2(D-2)$ Grassman variables $\xi_\alpha$ each of which is paired
either with an $\omega$, or with an explicit factor
$R^{-(2-\eta)}$. Since $\omega$ is itself bounded by $R^{-(2-\eta)}$
(\ref{omegbound}), we 
find
\bea 
\absval{\C P^r_{D,G}(x,R)}&<&\frac{R^{-2(D-2)(2-\eta)}}{(2(D-2)-k)!}
\sum_{P} \Bigg\vert \int d\phi  d\chi
\left(\frac{\phi^1\cdots\phi^m}{R^{m(2-\eta)}}
\frac{\chi^1\cdots\chi^{n+2(D-2)-k}}{R^{n(2-\eta)}}\right)\nn\\
&&\times \exp\left(\Tr \phi \comm{y}{\phi} 
+\Tr \phi\comm{\omega}{\chi}+\Tr \chi \comm{\omega}{\phi} 
+ \Tr \chi\comm{x}{\chi}\right)\Bigg\vert
\eea
where $P$ indicates all the possible permutations of indices that can
be generated.
The next step is to expand the $\phi\omega\chi$ terms to get
\bea 
\absval{\C P^r_{D,G}(x,R)}&<&R^{-2(D-2)(2-\eta)}
\sum_{P}\sum_l\frac{2^l}{l! (2(D-2)-k)!}\nn\\
&&\times \absval{\int d\phi
\left(\frac{\phi^1\cdots\phi^{m+l}}{R^{(m+l)(2-\eta)}}
\right)\exp\left(\Tr \phi \comm{y}{\phi}\right) }\\
&&\times\max_x\absval{\int d\chi\left(\frac{\chi^1\cdots
\chi^{n+2(D-2)-k+l}}{R^{n(2-\eta)}}\right)
\exp\left( \Tr \chi\comm{x}{\chi}\right)}.\nn
\eea
Finally, integrate out the $\chi$ fermions and use the fact that $x$
is bounded to obtain
\beq
\absval{ \C P^r_{D,G}(x,R)} < R^{-(2-\eta)2(D-2)} \sum_{r'} C_{r'}
\absval{ \C P^{r'}_{D,G'}(y,R) }\label{Pbound}
\eeq
where the $C_{r'}$ are constants. In the spirit of the notation
\ref{simplepfaff}, we have suppressed sums over the many possible
combinations of indices.

\noindent
Inserting the bound \ref{Pbound} into \ref{PIdef} gives
\beq\label{bndx}
\C I^r_{D, G}(R) < R^{-(2-\eta)2(D-2)} \sum_{r'} C_{r'} \int_{\C R_1} \prod_{\nu=1}^D dx_\nu \,
\absval{\Pf^{r'}_{D,G'} 
(y,R)} \deltafn{\left(1-\Tr x_\mu x_\mu\right)}
\eeq 
and we can now follow the bosonic procedure and integrate out the
$\omega$ and $\rho$  degrees 
 of freedom \pagebreak[4] to obtain
\bea
\C I^r_{D, G}(R)& <& R^{-(2-\eta)2(D-2)}
R^{-(2-\eta)(D-1)(g-g'-1)} 
\nonumber \\ 
&&\times \sum_{r'} C_{r'} \int_{\C R_1(G')} \prod_{\nu=1}^D dy_\nu \, 
\absval{\Pf^{r'}_{D,G'} 
(y,R)} \thetafn{\left(1-\Tr y_\mu y_\mu \right)}.
\eea
As before, replace $\thetafn{ ( 1 -\Tr y_\mu y_\mu )} = \int_0^1 dt
\deltafn{ (t-y_\mu y_\mu)}$ and rescale $t=[u/R]^{2-\eta}$ and $y_\mu
= \tilde{y}_\mu [u/R]^{1-\eta /2}$ giving
\bea
\C I^r_{D, G}(R)& <& R^{-(2-\eta)2(D-2)} R^{-(2-\eta)(D-1)(g-g'-1)}
\nonumber \\
&&\times \sum_{r'} C_{r'} \int_0^R {du \over u} [u/R]^
{(2-\eta)[(D-1)g' + 3r'/2]} \nonumber \\
&& \times \int_{\C R_1(G')} \prod_{\nu=1}^D
d\tilde{y}_\nu \,  
\absval{\Pf^{r'}_{D,G'} 
(\tilde{y},u)} \deltafn{\left(1-\Tr \tilde{y}_\mu \tilde{y}_\mu \right)}.
\eea
Since $u/R <1$, this can be reduced to
\beq\label{F17}
\C I^r_{D, G}(R)< R^{-(2-\eta)[2(D-2)+(D-1)(g-1)]}
\sum_{r'} C_{r'} \int_0^R {du \over u} u^
{(2-\eta)(D-1)g'} \C I^{r'}_{D,G'}.
\eeq 
We argue by induction, so assume that, for $G'$
\beq\label{F18}
\int_0^\infty dR R^{Dg'-1} R^{(D-2)g'} \C I^r_{D,G'}(R)
\eeq
converges for $D > 3$, and all choices of $r$. Then \ref{F17} gives
\beq\label{finalbnd}
\C I^r_{D, G}(R)< C R^{-(2-\eta)[2(D-2)+(D-1)(g-1)]} \, , \;\;\; R>1
\eeq
and in particular, by the usual power counting argument,  the
induction statement is true also for $G$. It 
remains to check that the induction statement is true for the smallest
possible regularly embedded subalgebra, which is $su(2)$. Since
$su(2)$ has no regularly embedded subalgebra, we can repeat the above
arguments with $\C G' =0$ and find
\beq
\C I^r_{D, SU(2)}(R)< C R^{-(2-\eta)[2(D-2)+2(D-1)]}
\eeq
so that \ref{F18} indeed converges for $D>3$.

Taking now $r=0$, we have discovered that, for any compact semi-simple
group $G$, 
\beq
\C I_{D, G}(R)< C R^{-(2-\eta)[2(D-2)+(D-1)(g-1)]}
\eeq
and in particular, the partition function $\C Z_{D,G}$ converges for
$D > 3$. For the correlation function \ref{correl} to converge, we
require
\beq
Dg + (D-2)g +k < 2[2(D-2)+(D-1)(g-1)]
\eeq
and so the critical value is
\beq
k_c = 2(D-3)
\eeq
independently of the gauge group. 

\section{Discussion}
\subsection*{Bosonic Theory}
For the bosonic theories, we have shown that the partition function
converges when $D \geq D_c$ and calculated $D_c$ for each of the
compact simple groups: 
\beq
\nonumber
\begin{array}{ccl}
D_c=5,&\;\;\;\;\;&SU(2)\\
D_c=4,&&SU(3)\\
D_c=3,&&\hbox{all other simple groups.}
\end{array}
\eeq
It is a simple exercise to extend the result to
the compact semi-simple groups since they are built out of the simple
groups. For example, $so(4)= su(2) \oplus su(2)$, so $\C Z_{D,SO(4)}$
converges when $D \geq D_c=5$. In addition, we have calculated the
critical degree $k_c$ for correlation functions, such that $\langle \C
C_k \rangle$ converges when $k<k_c$. Conversely, we have shown that
there always 
exists a correlation function of degree $k_c$ which diverges.

Restricting ourselves to $D>2$, it seems rather mysterious that the
only divergent partition functions occur for
$SU(2)$ with $D=3, 4$, and for $SU(3)$ with $D=3$. However, there is
an argument which quickly allows us to see why this is so.
Begin with the bosonic integral
\ref{bospartn}, and follow the usual procedure to  move
$X_1$ into the Cartan 
subalgebra and pick up the Weyl measure (\ref{weyl})
\beq
\C Z_{D,G} = A_1 \int \prod_{i=1}^l dX_1^i \Delta^2_G(X_1^i) \int
\prod_{\nu=2}^D dX_\nu 
e^{-S(X)}.
\eeq
We can expand the $X_\nu$ in terms of the basis
\ref{B11}
\beq
X_\nu = X_\nu^i H^i + X_\nu^\alpha E^\alpha \, , \;\;\; \nu=2 , \cdots
, D,
\eeq
and then change variables from the $X_\nu^\alpha$ to
$(D-1)(g-l)$ dimensional polar coordinates with radial variable $\varrho$
and angular variables $\{ \theta_a \}$. Then
\beq
\C Z_{D,G} = A_2 \int \left( \prod_{\mu=1}^D \prod_{i=1}^l dX_\mu^i \right)
\Delta^2_G(X_1^i) \int d \varrho \varrho^{(D-1)(g-l)-1} d\Omega
\exp (-S),
\eeq
and the action can be expanded
\beq
S(X) = \varrho^2 X_\mu^i X_\nu^j Q_{\mu \nu}^{ij}(\theta_a)  +
\varrho^3 X_\mu^i F_\mu^i (\theta_a) + \varrho^4 F_1 (\theta_a)
\eeq
where $Q$, $F$ and $F_1$ are some functions of the angles
$\theta_a$. The action is quadratic  in the $X_\mu^i$, so we can
integrate them out and find
\beq
\C Z_{D,G} = A_3  \int_0^\infty d \varrho \varrho^{(D-1)(g-l)-1}
\varrho^{-Dl -(g-l)}  \int d\Omega F_2( \theta_a ) \exp ( -\varrho^4
F_3(\theta_a) )
\eeq
where $F_2$ and $F_3$ are some functions of the $\theta_a$, and
certainly  $F_2$
is positive semi-definite. This integral diverges at
$\varrho =0$ when 
\beq
\label{nice}
D \leq {2(g-l) \over g-2l}.
\eeq
Any group satisfying \ref{nice} must have a divergent partition
function, and so this gives a quick and illuminating way of seeing
that for $SU(2)$ with $D=3$ and $4$, and for $SU(3)$ with $D=3$, the
partition functions are divergent.

\subsection*{Supersymmetric Theory}
In the supersymmetric case, we have shown that the partition function
converges in $D=4,6$ and $10$ with any compact semi-simple gauge
group, and that correlation functions of degree
\beq
k<k_c =2(D-3)
\eeq
are convergent independent of the gauge group. In the case of
$SU(r+1)$, this result corresponds to the conjecture of
\cite{Krauth:1999qw} based on Monte Carlo evaluation of the integrals
for small  $r$.

\chapter{The Supercharge}
\label{chap:coho}
We now turn our attention specifically to the supersymmetric
Yang-Mills matrix models, and address the question of which quantities
are invariant under the supercharges. 
As we shall see, the supercharges take on a particularly simple form
if we reformulate the theory as a cohomological matrix
model. We shall
give a brief introduction to these models in section \ref{begin}, and
then prove our result for these models in the following
sections. Finally,  in section \ref{equiv},
we show that the result can also be applied to the Yang-Mills matrix
theories.

\section{Introduction to Cohomological Matrix Models}
\label{begin}
A deep relation between the Yang-Mills matrix models and so called
\begin{it}cohomological\end{it} models was  
uncovered by Moore, Nekrasov and Shatashvili. In a remarkable paper
\cite{Moore:1998et},  they were able to predict the value of the
Yang-Mills 
partition function by using the cohomological theory. We 
shall discuss this in detail in chapter \ref{chap:3}
but, for now, give a brief introduction to the cohomological model.
To illustrate, we shall consider the $D=4$ model,
although the techniques which follow can be applied immediately to
the cases of six and ten dimensions by following \cite{Moore:1998et}. The
action\footnote{Note that we have changed the notation  from the
previous chapters  in order to agree
with the literature on this subject. We have used ``$\lambda$'' for the
fermions, since the Greek letter $\psi$ will
shortly be introduced for a slightly different purpose.} is
\beq\label{ymaction}
S_{Y \! M}= - \Tr \left( \quarter
[X_{\mu},X_{\nu}]^{2} + 
\overline{\lambda} \overline{\sigma}^{\mu}
[X_{\mu},\lambda] \right).
\eeq
In this chapter we shall mainly be concerned with the gauge group
$G=SU(N)$ model so that 
all fields are $\NbyN$ 
matrices. The
gauge fields
$X_{\mu}$ ($\mu = 1,\cdots, 4$) are restricted to the Lie algebra of
$G$ which is the set of traceless
hermitian matrices. The fermions $\lambda$, which are in the Weyl
representation, are complex traceless
Grassman matrices.  We follow \cite{src:wessbagg} to give an explicit
representation 
$\overline{\sigma}^{\mu}$ for the
$D=4$ Dirac matrices projected to a Weyl representation. Define
$\sigma^{i}$ to be the 
Pauli matrices
\beq
\begin{array}{ccc}
{\sigma^{1}= 
\left( \begin{array}{rr} 0 & 1 \\[3mm] 1 & 0 \end{array} \right) } &
\sigma^{2}= 
\left( \begin{array}{rr} 0 & -i \\[3mm] i & 0 \end{array} \right) &
\sigma^{3}=
\left( \begin{array}{rr} 1 & 0 \\[3mm] 0 & -1 \end{array} \right)
\end{array}
\eeq
and 
$\sigma^{4}= -i \, 1_{2}$. Then the $\overline{\sigma}^{\mu}$ are defined by
\beq\label{sigmabar}
\overline{\sigma}^{\mu \dot{a} a} = \epsilon^{\dot{a} \dot{b}}
\epsilon^{a b} \sigma ^{\mu}_{b \dot{b}}
\eeq
and we define
\beq
\overline{\lambda} = -i\sigma_{4} \lambda^{\dagger} =
-\lambda^{\dagger}.
\eeq
The partition function can be written
\beq 
\label{eq:I}
\C Z_{4,N} =
\int dXd\lambda dD \exp  \Tr \left( \quarter
[X_{\mu},X_{\nu}]^{2} + 
\overline{\lambda} \overline{\sigma}^{\mu}
[X_{\mu},\lambda] - 2 D^{2}
\right)
\eeq
where an auxiliary field $D$ with appropriate integration measure has been
added. The auxiliary field allows us to write down the
supersymmetry of this model in a nice linear form,
\beq
\begin{array}{lcl}\label{SS}
\delta_{\xi} X^{\mu} & = & -i \overline{\lambda} \overline{\sigma}^{\mu}
\xi +i \overline{\xi} \overline{\sigma}^{\mu} \lambda \\
\delta_{\xi} \lambda & = & i \sigma^{\mu \nu} \xi [X^{\mu},X^{\nu}] -2 \xi
D \\
\delta_{\xi} D & = & \half [X^{\mu},\overline{\lambda}] \overline{\sigma}^{\mu}
\xi + \half \overline{\xi} \overline{\sigma}^{\mu} [X^{\mu},\lambda]
\end{array}
\eeq
where $\sigma^{\mu \nu} = {1 \over 4} (\sigma^\mu \overline{\sigma}^\nu
-\sigma^\nu \overline{\sigma}^\mu)$. This formula was obtained simply
by taking the supersymmetry transformation laws for the four
dimensional Yang-Mills theory (for example from \cite{src:wessbagg})
and dimensionally reducing to zero dimensions. 

It is important to note
that the expression \ref{SS} is rather formal. If
our original $D=4$ space were Minkowski, then we would have 
$\overline{\sigma}^4 =1_2$. However, since we are working with a Euclidean
metric, in fact $\overline{\sigma}^4 =-i 1_2$. This means that, whilst
$\overline{\sigma}^i$ $(i=1,2,3)$ are hermitian,  $\overline{\sigma}^4$
is antihermitian. Then $\delta_{\xi}$ does not preserve hermiticity of
the matrix fields so that, for example, $\delta_\xi X^4$ is
antihermitian rather than 
hermitian. Nevertheless, the transformation \ref{SS} is formally
a symmetry of the action, and we shall make rigorous use of this in
chapter \ref{chap:3}.

The approach of \cite{Moore:1998et} is to make the following field
replacements.
First rewrite the fermions in terms of their hermitian and antihermitian
parts
\beq\label{splitferm}
\begin{array}{ccc}
\lambda_{1} & = & (\eta_{2} +i \eta_{1}) \\
\lambda_{2} & = & (\psi_{1} +i \psi_{2}). \\
\end{array}
\eeq
We also re-write the auxiliary field
\beq
D = H + \half [X_{1},X_{2}]
\eeq
By the usual contour shifting argument for a Gaussian integral, $H$
can be taken hermitian. This is true for the partition function
\ref{eq:I} and also the correlation functions \ref{intro:correln} and
so makes sense throughout the theory\footnote{For a more detailed
explanation of the contour shifting argument, we refer ahead to
equation  \ref{Hshift}.}. 
To obtain the related cohomological action, make the replacement
\beq
\label{phidef}
\begin{array}{ccr} 
\phi & = & \half (X_{3} + iX_{4}) \\
\phibar & = & -\half (X_{3} - iX_{4})
\end{array}
\eeq
and take $\phibar$ hermitian and $\phi$ antihermitian. This gives
\beq
\label{eq:Scoh}
\begin{array}{rl}
S_{E}^{0} \rightarrow S_{\rm{coh}} =
\Tr \, {(} & \!\!\!\!
H^{2} + H[X_{1},X_{2}] -\epsilon^{a b} \eta_{1}
[\psi_{a} ,X_{b}]
+\eta_{2} [\psi_{a} ,X_{a} ] \\ & \!\! - \eta_{a} [\phi ,
\eta_{a}] - \psi_{a} 
[\phibar , \psi_{a}] +[X_{a},\phi] [X_{a},\phibar] + [\phi,
\phibar]^{2} \; {)}
\end{array}
\eeq
The key point is that $\phi$ and $\phibar$ are taken to be
\begin{it}independent\end{it}. This is clearly not true in the
Yang-Mills model, and so we have defined an entirely new theory.

The supersymmetry \ref{SS} depends on two complex Grassman parameters
$\xi_a$, and so one can  break it down to four linearly
independent real supercharges.
One can easily write four linearly independent supercharges of
$S_{Y \! M}$ in terms of the new variables, and one of these is
\beq
\label{eq:delta}
\begin{array}{ll}
\delta X_{a} =  \psi_{a} &
\delta \psi_{a}  =  [\phi,X_{a}] \\
\delta \phibar  =  -\eta_{2} & \delta \eta_{2}  =  
-[\phi ,\phibar ] \\
\delta \eta_{1}  =  H & \delta H  =  [\phi , \eta_{1} ] \\
\delta \phi  =  0. 
\end{array}
\eeq
Since $\delta^{2}  = [\phi , \;\;]$, $\delta$ is nilpotent on gauge
invariant quantities. For interest, a representation of all four
supercharges is given in appendix \ref{app:super}, together with some
relations between them.

The action is $\delta$-exact. $S= \delta Q$, where
\beq
Q = \Tr (\eta_{1} [X_{1},X_{2}] +  \eta_{1} H -\psi_{a}
[X_{a},\phibar ] - \eta_{2} [\phi , \phibar ] )
\eeq
as can readily be checked. So the symmetry $\delta S =0$ is
manifest. The term cohomological  to describe the
theory is arrived at by analogy of $\delta$ with an exterior
derivative. 

In this chapter, we study the supersymmetry operator $\delta$, and in
particular we address the question of which quantities are
supersymmetric under $\delta$. Gauge invariant quantities are formed
from traces, and so we seek the general solution to the equation
\beq
\delta \Tr P = 0
\eeq
where $P$ is a polynomial in the matrix fields. In the analogy of
$\delta$ with an exterior derivative, this is the question of finding
the cohomology. An important example of
the use is to find the possible supersymmetric deformations of a
given action. One usually requires a result valid for any gauge group
SU(N), so we shall allow ourselves to make the assumption that $N$ is
suitably 
large. We shall show that
\beq	
\delta \Tr P = 0 \Leftrightarrow \Tr P = \delta \Tr Q + \Tr R( \phi)
\eeq
as long as the  degree of $P$ is less than ${2N \over 3}$.

The proof requires a number of steps. We  form a vector space
from the polynomials, and deal with issues of linear dependence in
section \ref{sec:polys}. A major technical difficulty  is that linearly
independent polynomials become dependent after applying a
trace. This is overcome in section \ref{sec:trace} by forming a suitable
quotient space. Then in section \ref{sec:d} the result is proved for a
simplified version of $\delta$ in which the $[\phi , \;\;]$
terms are absent. Finally, in section \ref{sec:final}, the strands are
drawn together to prove the result.

\section{Polynomials}
\label{sec:polys}

We wish to form a vector space from the polynomials, and eventually
argue by induction on  degree. However, 
there is a technical difficulty. Two polynomials which
\begin{it} look \end{it}
different, because they contain different strings of matrices
multiplied together, can turn out to be identical. At this stage, let
us be definite and make some careful definitions.
\begin{description}
\item[String]
A string of length $l$ is a map from $\{1,\cdots ,l\}$ into the set of
matrix fields. 
For example, a typical string of length $5$ might be 
\beq
\begin{array}{cccccc}
s = & X_{1} & \eta_{2} & \phibar & X_{1} & \phi \\
    & (1)   & (2)      & (3)     & (4)   & (5)
\end{array}
\eeq
\item[Monomial]
A monomial of degree $d>0$ is the matrix product of $d$ matrix fields. The
monomial of degree $0$ is the identity matrix.
For example, a typical monomial of degree $5$ might be
\beq
m =  X_{1} \, \cdot \, \eta_{2} \, \cdot \, \phibar \, \cdot \, X_{1}
\, \cdot \, \phi  
\eeq
where $\, \cdot \,$ indicates matrix multiplication.

Each string of length $l$ is naturally associated to a
monomial of degree $l$ by applying matrix multiplication
between adjacent fields in the string.
\item[Polynomial]
A polynomial of degree d is a linear combination of a finite number
of monomials whose highest degree is d.
\end{description}

One can form an abstract vector space $V_{s}$ over $\mathbb{C}$ by taking the
strings as the basis. In $V_{s}$, the strings are linearly
independent. However, as polynomials, the strings are not
necessarily linearly independent. This is most easily seen when the
matrix size $N$ is $1$ so that bosonic matrices commute. Then the two
independent strings $X_{1} \; X_{2}$ and $X_{2} \; X_{1}$ are
identical as polynomials. Even when $N>1$ so that matrices do not
commute, it is possible for independent strings to be linearly
dependent as polynomials. A trivial example is that $\psi^{N^{2}}
\equiv 0$ when $\psi$ is a traceless hermitian $\NbyN$ fermion.

This problem can be overcome by considering only polynomials of degree
smaller than the matrix size. Assume that the matrix fields are $\NbyN$ and
hermitian. They may also have the constraint of tracelessness, but no
other constraints. Then the strings of length less than $N$ are
linearly independent as polynomials.

To see this, 
denote the strings of length less than $N$ by $\{ s^{b} \}$ and the
corresponding 
monomials $\{ m^{b} \} $.

\noindent
Suppose 
\beq
\label{eq:ld}
\lambda^{b} m^{b} \equiv 0
\eeq
for some $\lambda^{b} \in
\mathbb{C}$ and (without loss of generality) $\lambda^{1} \neq 0$.
Write $m^{1} = Y^{1} \cdots Y^{d}$ where the $Y^{i}$ are matrix fields
and the degree of $m^{1}$ is $d<N$.
Then, in particular, the term
$Y_{12}^{1} Y_{23}^{2} \cdots Y_{d,d+1}^{d}$ is absent from
\ref{eq:ld}. But the only monomial which gives rise to this term is
$m^{1} = Y^{1} \cdots Y^{d}$. Therefore $\lambda^{1}=0$, and this is a
contradiction.

Note that  no assumptions are made about which of the matrix fields are
fermionic and which bosonic.

\section{The trace}
\label{sec:trace}
It will be convenient  to work with  polynomials rather than traces
of polynomials. Unfortunately, two independent polynomials can have
identical trace. Defining an equivalence relation $P \sim Q
\Leftrightarrow \Tr P = \Tr Q$, we would like to form the
quotient space $V_{p}/ \! \! \sim$, where $V_{p}$ is the vector space of
polynomials.

Consider a polynomial $P(A^{a})$, where $ \{ A^{a}: a=1,\cdots ,M \} $
are the matrix fields. Assume that the only constraints
which may be applied to the fields are hermiticity and
tracelessness. Define an ordering $\mathcal{O}$
such that
\begin{itemize}
\item $\mathcal{O}$ acts individually on each monomial term in $P$
\item $\mathcal{O}$ cyclically permutes each monomial in $P$ into a
preferred form with a sign to respect fermion statistics
\end{itemize}
An example of such an ordering would be to define
$A^{1}>A^{2}>A^{3}>\cdots$. In this case, for example, $\mathcal{O}(A^{4}A^{2}A^{5}) =
(-1)^{F_{4}(F_{2}+F_{5})} A^{2}A^{5}A^{4}$ where $F_{a}$ is the
fermion number of $A^{a}$.

\noindent
Then, for $\degree (P) <N$
\beq
\label{eq:order}
\Tr P = 0 \Leftrightarrow \mathcal{O}(P) = 0
\eeq
so that $\mathcal{O}$ gives a mapping to the quotient space.

To see that \ref{eq:order} is true, first note that
$\mathcal{O} P =0 \Rightarrow \Tr \mathcal{O} P =0 \Rightarrow \Tr P
=0$ by the cyclic property of trace.

\noindent
Conversely, suppose $\mathcal{O} P \neq 0$. Consider a particular
monomial term in $\mathcal{O} P$:
\beq
\mathcal{O} P = \lambda Y^{1} \cdots Y^{M} + \cdots
\eeq
where  $\lambda$ is some non-zero coefficient.
Then $\Tr P = \Tr \mathcal{O} P$ contains the term:
\beq
\Tr P = \lambda Y_{12}^{1} Y_{23}^{2} \cdots Y_{M1}^{M} + \cdots
\eeq
Since $\mathcal{O} P$ is ordered, the only monomial that can give such
a term is $Y^{1} Y^{2} \cdots Y^{M}$. Therefore this term cannot be
cancelled and so $\Tr P \neq 0$.
Deduce $\Tr P =0 \Rightarrow \mathcal{O} P = 0$. 

An ordering operator $\mathcal{O}$ is not the most useful way of
dealing with the trace. Since there is no way that $\mathcal{O}$
will commute with any form of supersymmetry operator, it is more helpful to
use the following:

\noindent
Let $P_{M}$ be a polynomial in which all of the terms are of degree
$M$. Then for $M<N$
\beq
\label{eq:cycsum}
\Tr P_{M} = 0 \Leftrightarrow
\sum_{\stackrel{\rm{cyclic}}{\rm{perms}}} P_{M} =0
\eeq
The cyclic permutations act on the matrix fields in the monomials, and
include a sign to respect fermion statistics. For example, suppose $P
= \lambda F_{1} F_{2} B_{3} + \mu B_{1} B_{2} B_{4}$ where the $F_{i}$ are
fermionic and the $B_{i}$ bosonic matrix fields. Then the cyclic
permutation $\sigma 
= (123)$ acts as
\beq
\sigma P = - \lambda F_{2} B_{3} F_{1} + \mu B_{2} B_{4} B_{1}
\eeq
Equation \ref{eq:cycsum} follows easily from \ref{eq:order}.
First note that \beq
\sum_{\stackrel{\rm{cyclic}}{\rm{perms}}} P_{M} =
\sum_{\stackrel{\rm{cyclic}}{\rm{perms}}} \mathcal{O} P_{M}
\eeq
Then
\beq
\Tr P_{M} =0 \Rightarrow \mathcal{O} P_{M} = 0 \Rightarrow
\sum_{\stackrel{\rm{cyclic}}{\rm{perms}}} \mathcal{O} P_{M} =0
\Rightarrow \sum_{\stackrel{\rm{cyclic}}{\rm{perms}}} P_{M} =0
\eeq
Conversely
\beq
\Tr \sum_{\stackrel{\rm{cyclic}}{\rm{perms}}} P_{M} = M \Tr P_{M}
\eeq
and so
\beq
\sum_{\stackrel{\rm{cyclic}}{\rm{perms}}} P_{M} = 0 \Rightarrow \Tr
P_{M} = 0
\eeq

\section{Decomposition of the supercharge}
\label{sec:d}
If the expression for $\delta$ (\ref{eq:delta}) did not contain the
commutator terms, our task to classify the supersymmetric quantities
would be much simpler. In this section, we decompose the supercharge
into two parts, and prove our result for the simpler part. This will
allow us to tackle the full supercharge in the next section.

\noindent 
Returning to the specific theory under discussion, let us write
\beq
A^{1}=X_{1}, \;\; A^{2}=X_{2},\;\; A^{3}=\phibar ,\;\; A^{4}=\eta_{1}
\eeq
\beq
B^{1}=\psi_{1},\;\; B^{2}=\psi_{2},\;\; B^{3}=-\eta_{2} ,\;\; B^{4}=H
\eeq
Then the supersymmetry $\delta$ can be written as
\beq
\label{eq:genericdelta}
\begin{array}{lcc}
\delta A^{i} = B^{i} & , & \delta B^{i} = [\phi ,A^{i}] \\
\delta \phi = 0
\end{array}
\eeq 
If one considers also the six- and ten-dimensional cohomological matrix
models, one finds an identical form for $\delta$ \cite{Moore:1998et},
and so the results from this point on apply equally to all three theories.

\noindent
Define two new operators $d$, $\Delta$ by
\beq\label{d}
\begin{array}{lcc}
d A^{i} = B^{i} & , & d B^{i} = 0 \\
d \phi = 0
\end{array}
\eeq
\beq
\begin{array}{lcc}
\Delta A^{i} = 0 & , & \Delta B^{i} = [\phi ,A^{i}] \\
\Delta \phi = 0
\end{array}
\eeq 
One can very easily check that the following relations hold on
polynomials
\beq
\begin{array}{lcl}
i) & \hspace{1cm}& \delta^{2} = [\phi ,\;\;] \\
ii) &&  d^{2} = 0 \\
iii) & & \Delta^{2} = 0 \\
iv) & & \delta = d + \Delta \\
v) & & \{ d,\Delta \} = \delta^{2} = [\phi ,\;\;]
\end{array}
\eeq
and that the operator $d$ has the useful property
\beq
\label{eq:dcomm}
d \sum_{\stackrel{\rm{cyclic}}{\rm{perms}}} P_{M} =
\sum_{\stackrel{\rm{cyclic}}{\rm{perms}}} d P_{M}
\eeq
which will allow us to deal with the
trace.

This nice property \ref{eq:dcomm} makes the operator $d$ much simpler
to deal with. We begin by forgetting the trace and proving  the
following result for a 
polynomial. Let $P$ be a 
polynomial of degree less than $N$, and suppose $dP=0$. Then  

\beq
\label{eq:dpoly}
d P=0 \Rightarrow P=dQ+R(\phi )
\eeq 
for some polynomials $Q$ and $R$.

To show \ref{eq:dpoly}, we use
induction on the degree of $P$.
The case $\degree (P)=0$ is simple since then $P = \lambda I = R(\phi )$.
When $\degree (P)>0$, expand
\beq
\label{eq:d0}
P = A^{i} S^{i} + B^{i} T^{i} + \phi U + \lambda I
\eeq
for some polynomials $S^{i}$, $T^{i}$ and $U$, and a constant $\lambda$. 
Applying $d$,
\beq
0 = d P = B^{i} S^{i} + (-1)^{A^{i}} A^{i} d S^{i} + (-1)^{B^{i}}
B^{i} d T^{i} + \phi d U
\eeq
where the notation $(-1)^{A^{i}}$ is shorthand for $\pm 1$ respectively as
$A^{i}$ is bosonic or fermionic. Then in particular, since $d$ maps
bosons to fermions, $(-1)^{B^{i}} = (-1)^{A^{i}+1}$.

\noindent
Since $\degree (P)<N$, the strings are linearly independent (section
\ref{sec:polys}), and we deduce 
\beq
\label{eq:d1}
S^{i} + (-1)^{A^{i}+1}d T^{i} = 0
\eeq
\beq
\label{eq:d2}
d S^{i} = 0
\eeq
\beq
\label{eq:d3}
d U = 0.
\eeq
(and note that since $d^{2}=0$, \ref{eq:d2} is implied by \ref{eq:d1}). 

\noindent
By induction, $dU=0$ implies
\beq
\label{eq:d4}
U = d V + W(\phi )
\eeq
where $V$ and $W$ are polynomials.
Then substituting \ref{eq:d1} and \ref{eq:d4} back into \ref{eq:d0},
\beq
\begin{array}{ccl}
P &=& A^{i} (-1)^{A^{i}} d T^{i} + B^{i} T^{i} + \phi \left( dV +
W(\phi) \right) +\lambda I \\
&=& d(A^{i} T^{i} + \phi V ) + \phi W(\phi ) + \lambda I \\
&=& d Q + R(\phi)
\end{array}
\eeq
and the result \ref{eq:dpoly} follows.

Finally in this section, we introduce the trace.
Let $P$ be a polynomial of degree less than $N$, and suppose $d \Tr P
= 0$. Then 
\beq
\label{eq:dtrace}
d \Tr P = 0 \Rightarrow \Tr P = d \Tr Q + \Tr R(\phi )
\eeq
for some polynomials $Q$
and $R$.

To see this, write $P=P_{0} + \cdots + P_{M}$ where each $P_{i}$
contains only monomials 
of degree $i$.
Then, since $d$ preserves the degree of monomials,
\beq
\begin{array}{ccl}
d \Tr P = 0 &  \Rightarrow & d \Tr P_{i} = 0 \\
            &  \Rightarrow & \Tr d P_{i} = 0 \;\;\;\;(i=0,\cdots,M).
\end{array}
\eeq
The case of $i=0$ is simple. For $i>0$,  using \ref{eq:cycsum} implies
\beq
\sum_{\stackrel{\rm{cyclic}}{\rm{perms}}} d P_{i} =0
\eeq
and using \ref{eq:dcomm} 
\beq
d \sum_{\stackrel{\rm{cyclic}}{\rm{perms}}} P_{i} =0.
\eeq
Then \ref{eq:dpoly} gives
\beq
\sum_{\stackrel{\rm{cyclic}}{\rm{perms}}} P_{i} = d Q_{i} + R_{i}(\phi
)
\eeq
for some polynomials $Q_{i}$ and $R_{i}$, so that
\beq
\Tr P_{i} = {1 \over i}\Tr ( d Q_{i} + R_{i}(\phi ) )
\eeq
by the cyclic property of trace.
Then summing over $i$ gives the result.

\section{Extension to the full supercharge}
\label{sec:final}
The task now is to extend the result from $d$ to $\delta$. The
commutator terms in the definition of $\delta$ make
it much harder to deal with the trace. Specifically, $\delta$ does not
commute with the sum over cyclic permutations. Instead,
we proceed 
with a less direct approach, and make use of the result
for $d$.

\noindent
We begin with a technical result.
Suppose $P_{k}$ is a polynomial of degree $k$ satisfying 
$d \Delta \Tr P_{k}=0$. Then there exists a polynomial $T$ such that
\beq\label{tech}
d \Delta \Tr P_{k}=0 \Rightarrow
d \Tr P_{k} = (d + \Delta ) \Tr T,
\eeq
as long as $N> {3k \over 2}$.

\noindent
The proof follows an inductive argument. By \ref{eq:dtrace},
\beq
d \Delta \Tr P_{k} = 0 \;\;\;\; \Rightarrow \;\;\;\; \Delta \Tr P_{k}
= - d \Tr P_{k+1} + R_{k+1}(\phi)
\eeq
for some polynomials $P_{k+1}$ and $R_{k+1}$ of degree $k+1$. Since
neither $d$ nor $\Delta$ can produce monomials only in $\phi$,
$R_{k+1}(\phi ) = 0$. Then
\beq
d \Tr P_{k} = (d + \Delta ) \Tr P_{k} + d \Tr P_{k+1}
\eeq
On any monomial, $d$ acts to increase the number of fields of type
$B^{i}$ by $1$, 
whilst $\Delta$ acts to decrease the number of $B^{i}$ by $1$.

Let $M_{k}$ be the maximum number of $B^{i}$ occurring in any term of
$P_{k}$. Then since $\Delta \Tr P_{k} = d \Tr P_{k+1}$, we have
\beq
M_{k+1} = M_{k} - 2
\eeq
Proceed inductively to find
\beq
d \Tr P_{k} = (d+\Delta ) \Tr ( P_{k} + P_{k+1} + \cdots + P_{k+q})
+ d \Tr P_{k+q+1}
\eeq
where $P_{k+q+1}$ contains no $B^{i}$ fields at all. Then $\Delta \Tr
P_{k+q+1} = 0$ and so
\beq
d \Tr P_{k} = (d + \Delta ) \Tr ( P_{k} + \cdots + P_{k+q+1} )
\eeq
which proves the result as long as $N>k+q+1$ so that each inductive
step is valid. Noting that the case $M_{k}=k$ is special and can be
reduced to the case $M_{k}=k-1$, one finds that $N>{3k \over 2}$ is a
sufficient condition.

We are now ready to prove the main result of this chapter.
Suppose the matrix fields are of size $N$, and $P$ is a polynomial in
the matrix fields. Then for $\degree(P) < {2N \over 3}$,
\beq
\label{eq:cohom}
\delta \Tr P = 0 \Leftrightarrow \Tr P = \delta \Tr Q + \Tr R(\phi)
\eeq
where $Q$ and $R$ are polynomials.

To show this,
write $P= P_{0}+\cdots +P_{M}$ where $P_{i}$ contains monomials only
of degree $i$. Then
\beq
\delta \Tr P = 0 \;\;\;\; \Rightarrow \;\;\;\; (d+\Delta ) \Tr P = 0
\eeq
and since $d$ preserves degree whilst $\Delta$ increases degree by
$1$, we have
\beq
\begin{array}{l}
\Delta \Tr P_{M} = 0 \\
\Delta \Tr P_{i} + d \Tr P_{i+1} = 0,\;\;\;\; i=0,\cdots ,M-1 \\
d \Tr P_{0} = 0
\end{array}
\eeq
By \ref{eq:dtrace},
\beq
d \Tr P_{0} = 0 \;\;\;\; \Rightarrow \;\;\;\; \Tr P_{0} = d \Tr Q_{0}
+ \Tr R_{0}(\phi )
\eeq
and
\beq
\begin{array}{ccl}
\Delta \Tr P_{0} + d \Tr P_{1} = 0 & \Rightarrow & \Delta ( d \Tr
Q_{0} + \Tr R_{0}(\phi ) ) + d \Tr P_{1} = 0 \\
 & \Rightarrow & d ( - \Delta \Tr Q_{0} + \Tr P_{1} ) = 0 \\
 & \Rightarrow & \Tr P_{1} = \Delta \Tr Q_{0} + d \Tr Q_{1} + \Tr
R_{1} (\phi ).
\end{array}
\eeq
Repeating the same argument inductively gives
\beq
\Tr P_{i} = \Delta \Tr Q_{i-1} + d \Tr Q_{i} + \Tr R_{i}(\phi
),\;\;\;\; i=1,\cdots ,M
\eeq
for some polynomials $Q_{i}$ and $R_{i}(\phi )$, so that
\bea
\Tr P &=& d \Tr Q_{0} + \Tr R_{0}(\phi) + \sum_{i=1}^{M} \Delta \Tr
Q_{i-1} + d \Tr Q_{i} + \Tr R_{i}(\phi) \nonumber \\ \label{y}
&=& (d + \Delta ) \sum_{i=0}^{M-1} \Tr Q_{i} + \sum_{i=0}^{M} \Tr
R_{i}(\phi) + d \Tr Q_{M}.
\eea
If we now apply  $\delta$ to \ref{y} we find
\beq
\delta \Tr P = 0 \;\;\;\; \Rightarrow \;\;\;\; \Delta d \Tr Q_{M} = 0
\eeq
and so the technical result \ref{tech} gives
\beq\label{x}
d \Tr Q_{M} = (d + \Delta) \Tr S
\eeq
for some polynomial $S$. Then, since $\delta=d+\Delta$, substituting
\ref{x} back into \ref{y}   proves the result. 

\section{The Yang-Mills Supercharge}
\label{equiv}
The only non-trivial change in moving from the Yang-Mills model to the
Cohomological model was the change of variables (\ref{phidef})
\beq
\label{phidef2}
\begin{array}{ccr} 
\phi & = & \half (X_{3} + iX_{4}) \\
\phibar & = & -\half (X_{3} - iX_{4})
\end{array}
\eeq
since in the Cohomological formulation, we took $\phi$ and $\phibar$
independent and
respectively antihermitian and hermitian. Thus we can
write the supercharge \ref{eq:delta} in the Yang-Mills formulation
simply by making the replacement \ref{phidef2},
\beq
\label{ymcharge}
\begin{array}{ll}
\delta X_{a} =  \psi_{a} &
\delta \psi_{a}  =  [\half(X_3+iX_4),X_{a}] \\
\delta \eta_{1}  =  H & \delta H  =  [\half(X_3+iX_4) , \eta_{1} ] \\
\delta X_3  =  \eta_{2} & \delta \eta_{2}  =  
{i \over 2}[X_4 , X_3 ] \\
\delta X_4   =  i \eta_2 
\end{array}
\eeq
where now $X_3$ and $X_4$ are independent Hermitian
matrices\footnote{Here we refer the reader again to the note made
after equation \ref{SS}. The Yang-Mills supercharge \ref{ymcharge}
does not preserve hermiticity of the matrix fields since we are using
Euclidean metric.}. If we
label the matrix fields $A^a$, with $A^1=X_1$, $A^2=\psi_1$, etc, then
\ref{ymcharge} can be rewritten as a first order differential operator
\beq
\delta = (\delta A^a) {\partial \over \partial A^a}
\eeq
where we sum over the repeated index $a$ and, for example, $\delta A^1
= \delta X_1 = \psi_1$.  

In an identical way, we can write the supercharge of the cohomological
theory as a first order differential operator. Thus, the result of the
previous sections 
amounts to finding the general solution of the first order
differential equation
\beq
\label{diff}
\delta_{CoHo} f = 0
\eeq
with the constraint that the function of matrix fields $f$ is of the
form $f= \Tr P$ with $P$ a polynomial.

Lets now take a solution $f$ to this equation. 
Up to this point, $\phi$ and $\phibar$ have been respectively
antihermitian and hermitian matrix fields (and of course,
independent). However, since $f$ is a polynomial, it is a trivial
matter to analytically continue $f$ so that $\phi$ and $\phibar$ become
general complex matrices. Since this is an analytic continuation, in
particular the differential equation \ref{diff} still holds.

Now change variables from $\phi$ and $\phibar$ to $X_3$ and $X_4$
using equation \ref{phidef}. Then, using the chain rule, \ref{diff}
becomes
\beq
\delta_{Y \! M} f = 0
\eeq
where $\delta_{Y \! M}$ is of course precisely the differential
operator which 
is the Yang-Mills supercharge \ref{ymcharge}. Of course, $X_3$ and
$X_4$ are general complex 
matrices, but we can now restrict their domain to the hermitian
matrices to arrive back at the Yang-Mills theory.

Thus we have discovered that, if we have a solution
$\delta_{CoHo} \Tr P =0$
in the cohomological theory, then we can take $P$ and simply make the
replacement \ref{phidef2} to obtain a solution in the Yang-Mills theory.
Conversely, if we have a solution $\delta_{Y \! M} \Tr P =0$ in the
Yang-Mills theory, this gives us a solution in the cohomological
theory.

\noindent
Then the result of the previous section extends immediately to the
Yang-Mills theory,
\beq	
\delta_{Y \! M} \Tr P = 0 \Leftrightarrow \Tr P = \delta_{Y \! M}  \Tr
Q + \Tr R  (X_3 + i X_4 )
\eeq
as long as the  degree of $P$ is less than ${2N \over 3}$.

\section{Concluding Remarks}

We have considered the $SU(N)$ Yang-Mills and cohomological matrix
models in four, six and 
ten dimensions, and shown that
\bea
\label{eq:gen}
\delta_{CoHo} \Tr P = 0 &\Leftrightarrow& \Tr P = \delta_{CoHo} \Tr Q + \Tr R
( \phi)\\ \label{gen2}
\delta_{Y \! M} \Tr P = 0 &\Leftrightarrow& \Tr P = \delta_{Y \! M}  \Tr
Q + \Tr R  (X_3 + i X_4 )
\eea
as long as the  degree of $P$ is less than ${2N \over 3}$.

Although the large $N$ limit is a case of particular interest, it
would also be interesting to understand what happens when $N$ is
small, or the gauge group is not $SU(N)$. At present, we do not know of any
counter examples to the general formulae \ref{eq:gen}, \ref{gen2} in
these cases. 
It would also be interesting to understand whether the result
can be extended to a general gauge invariant quantity consisting of
an arbitrary function of traces.

\chapter{The Deformation Approach}
\label{chap:3}
We now return to the approach of Moore, Nekrasov and Shatashvili
\cite{Moore:1998et}. These authors begin with the cohomological action
\ref{eq:Scoh} with gauge group $SU(N)$, and consider the partition function
\beq
\int d\phi dX_1 dX_2 d \psi_1 d\psi_2 d\eta_2 d \eta_1 d\phibar dH
e^{- S_{coh} }. 
\eeq 
They add additional terms to the action $S_{coh}
\rightarrow S_{coh} + \epsilon \Delta S_{coh}$ in such a way as to
preserve some  
supersymmetry. They are then able to use Witten's localisation principle
\cite{Witten:1992xu} to 
integrate  out  the fields $H, \phibar, \eta, \psi, X$, leaving an
integral over just $\phi$, and use the gauge symmetry to diagonalise
$\phi$ 
in the usual way. The result is an integral of the form
\beq\label{mns}
\int d\phi_1 \cdots d\phi_{N-1} z(\phi_1, \cdots, \phi_{N-1})
\eeq
where the $\phi_i$ are the eigen-values of $\phi$. With the form of
$z$ which is obtained from these manipulations, the integral is
divergent. However, MNS complete the contours of integration in either
the upper or lower half plane following a certain prescription, and
perform the contour integrals. In dimensions $D=4,6,10$, the results
are identical to the value of the Yang-Mills partition function
\ref{intro:partn} in every case that it is known either numerically or
exactly. Furthermore, recently the result has been extended to some
groups other than $SU(N)$ and compared to numerical calculations, and
again the results agree \cite{Staudacher:2000gx,Krauth:2000bv}.

There are two puzzles in this calculation. The first is the question
of why it should be allowed to replace the Yang-Mills theory with the
cohomological theory, and the second is why following the MNS contour
prescription gives the correct result. It was hoped that finding the
answer to the first question would naturally answer the second (see
for example \cite{Hoppe:1999xg} in which these methods are applied to
the one-dimensional theory).

In this chapter, we carefully apply the deformation method of MNS
directly to the Yang-Mills model. This involves finding a deformation
of the Yang-Mills action for which we can be sure that the integrals
converge at every step. The final result is closely related to the
formula 
of MNS, and we indicate how the result as it stands comes to
be  divergent. However, sadly the MNS contour prescription does not
arrise naturally and we must again impose it by hand.

\section{Yang-Mills Integral}
We begin by recalling the model. We shall discuss the $D=4$, $SU(N)$
model in detail. 
The  action is (\ref{ymaction})
\beq\label{ymaction2}
S_{Y \! M}= - \Tr \left( \quarter
[X_{\mu},X_{\nu}]^{2} + 
\overline{\lambda} \overline{\sigma}^{\mu}
[X_{\mu},\lambda] \right)
\eeq
All fields are $\NbyN$ 
matrices and transform in the adjoint representation of the gauge
group $G = SU(N)$. The
gauge fields
$X_{\mu}$ ($\mu = 1,\cdots, 4$) are restricted to the Lie algebra of
$G$ which is the set of traceless
hermitian matrices. The fermions $\lambda_a$ $(a=1,2)$ are complex traceless
Grassman matrices. The $\overline{\sigma}^{\mu}$ are the $2 \! 
\times \! 2$ matrices  defined in \ref{sigmabar}. The
matrix integral 
giving the partition function is
\beq 
\label{eq:zzz}
\C Z =
\int dXd\lambda dD \exp  \Tr \left( \quarter
[X_{\mu},X_{\nu}]^{2} + 
\overline{\lambda} \overline{\sigma}^{\mu}
[X_{\mu},\lambda] - 2 D^{2}
\right)
\eeq
As usual, an auxilliary field $D$ 
has been 
added. So that the integral over $D$ does not affect the result, we
must fix the measure for $D$,
\beq\label{Dmeas}
\int dD e^{-2 \Tr D^{2}} =1.
\eeq
For completeness let us fix the integration measure of the other
fields now. Write any
hermitian matrix $Q$ in terms of its real and imaginary parts
\beq
Q = Q^{S} + i Q^{A}
\eeq
so that $Q^{S}$ and $Q^{A}$ are respectively symmetric and
antisymmetric  real $\NbyN$ matrices. We define
\beq
d Q = 2^{\mp \frac{1}{2}N(N-1)}\prod_{i\geq j}d Q^{S}_{i j}
\prod_{i>j}dQ^{A}_{i j}
\eeq
taking $-$ when $Q$ is bosonic, and $+$ when $Q$ is Grassmann. The
leading powers of $2$ may seem rather cumbersome, but they have an
advantage as we shall see in equations
\ref{eq:hermmeasure} and \ref{eq:hermmeasure2}. Since there are equal
numbers of bosons and 
fermions, these powers cancel in any case.

It is often inconvenient to integrate hermitian matrices directly, so
we define for a general complex matrix $M$
\beq\label{tilde1}
\widetilde{M} = {1 \over 2} (M + M^{T}) + {1 \over 2 i} (M-M^{T})
\label{eq:tilde}
\eeq
and conversely
\beq
M = {1\over 2}(\widetilde{M} +\widetilde{M}^{T}) + {i\over
2}(\widetilde{M}-\widetilde{M}^{T})
\label{eq:conversetilde}
\eeq
For $Q$ hermitian, observe
\beq
\widetilde{Q} = Q^{S} +Q^{A}
\eeq
so that $Q \leftrightarrow \widetilde{Q}$ gives a 1-1 correspondence
between the real and the hermitian $\NbyN$ matrices. With respect to
the new variables, the measure is
\beq
\label{eq:hermmeasure}
dQ = d\widetilde{Q}
\eeq
where
\beq\label{eq:hermmeasure2}
d\widetilde{Q} = \prod_{i,j}d\widetilde{Q}_{i j}
\eeq
is the natural measure on $\mathbb{R}^{N^{2}}$.

\label{sec:integrals}
This scheme has the advantage that $\Tr Q^{2} = \Tr \Qtil \Qtil^{T}$
so that a hermitian Gaussian integral is
\beq
\label{eq:gauss}
\int d Q e^{- \Tr Q^{2}} = \int d \Qtil
e^{-\sum_{i,j}(\Qtil_{ij})^{2}} = 
\pi^{N^{2}/2}
\eeq
To integrate over the traceless hermitian matrices, insert 
$\delta (\Tr Q )$.
One finds
\beq\label{tilde2}
\int_{\Tr Q = 0} d Q e^{- \Tr Q^{2}} = 
{\pi^{N^{2}-1 \over 2} \over \sqrt{N} }.
\eeq
A typical fermionic integral over $\chi$ and $\xi$ traceless hermitian
grassman matrices is
\beq\label{typferm}
\int_{\Tr \chi = \Tr \xi = 0} d\chi d\xi e^{i \Tr \chi \xi} =
\int_{\Tr \widetilde{\chi} = \Tr \widetilde{\xi} = 0}
d\widetilde{\chi} d\widetilde{\xi}  e^{i \Tr \widetilde{\chi}^{T}
  \widetilde{\xi}} 
= i^{N^{2}-1} N
\eeq
as can readily be checked.

The first step is to replace $D$ in \ref{eq:zzz} with
\beq\label{Hdef}
D = H+ \half \comm{X_1}{X_2}.
\eeq
Then the auxilliary part of the integral becomes
\beq\label{Hshift}
\int dH \exp \left\{ -2\Tr (H+ \half \comm{X_1}{X_2})^2 \right\} =
\int d \Htil \exp \left\{-2 
\sum_{i,j} ( \Htil_{ij} + i \Ctil_{ij})^2 \right\}
\eeq
where $C={-i \over 2} \comm{X_1}{X_2}$. The contour of integration of each
of the $\Htil_{ij}$ is displaced from the real axis by
$-i \Ctil_{ij}$. However, we can use the usual argument for Gaussian
integrals to shift the contours down onto the real axis. This means we
can take the $H$ defined in \ref{Hdef} to be hermitian.

At present, $H$ has the normalisation \ref{Dmeas} which is inherited
from $D$, whilst the $X_\mu$ have measure normalised by
\ref{tilde2}. For later convenience, we now  exchange the
normalisations of the measures of $X_4$ and $H$. This of course leaves
the matrix integral \ref{eq:zzz} unaffected. Thus the measure of $X_4$
is now normalised so that
\beq\label{X4norm}
\int dX_4
\deltafn{(\Tr X_4 )} \, e^{-2\Tr X_4^2}=1
\eeq
whilst $X_1$, $X_2$, $X_3$ and $H$ are all normalised according to
\ref{tilde2}.

We also follow the notation of the previous chapter (\ref{splitferm})
and split the 
fermions into
hermitian and antihermitian parts
\beq
\begin{array}{ccc}
\lambda_{1} & = & (\eta_{2} +i \eta_{1}) \\
\lambda_{2} & = & (\psi_{1} +i \psi_{2})
\end{array}
\eeq
so that one of the supercharges is (\ref{ymcharge})
\beq
\label{ymdelta}
\begin{array}{ll}
\delta X_{a} =  \psi_{a} &
\delta \psi_{a}  =  [\half(X_3+iX_4),X_{a}] \\
\delta \eta_{1}  =  H & \delta H  =  [\half(X_3+iX_4) , \eta_{1} ] \\
\delta X_3  =  \eta_{2} & \delta \eta_{2}  =  
{i \over 2}[X_4 , X_3 ] \\
\delta X_4   =  i \eta_2.
\end{array}
\eeq
We note that we can scale the exponent in \ref{eq:zzz} by a constant
and leave the integral invariant. Then, for later convenience, we
shall include include an 
extra  factor $\half$ so that, in terms of the new variables, the
action becomes
\bea
S &=& \Tr \left( (H + \half \comm{X_1}{X_2})^2 - {1 \over
4} \sum_{\mu>\nu}\comm{X_\mu}{X_\nu}^2 \right.  \\
&&
\left.  - \epsilon_{ab} \eta_1 \comm{\psi_a}{X_b}
-\eta_a \half \comm{ (X_3+iX_4)}{\eta_a} -\psi_a \half \comm{
(-X_3+iX_4)}{\psi_a} + \eta_2 \comm{\psi_a}{X_a} \right). \nonumber
\eea
The action is $\delta$-exact, $S= \delta \Tr Q$, where
\beq\label{Q}
Q = \left( \eta_{1} [X_{1},X_{2}] +  \eta_{1} H + \half \psi_{a}
 [X_{a},X_3-iX_4] - {i \over 2} \eta_{2} [X_3 , X_4 ] \right)
\eeq
as can readily be checked. So the symmetry $\delta S =0$ is
manifest.

\section{Deformation}
Our tactic for calculating the partition function is to add small mass
terms to the action in a prescribed way that will make the integrals
easy, and then send the masses to zero afterwards. Of course, we shall
eventually have to worry about whether the integration commutes with
the limit of 
masses going to zero, but for the moment we concentrate on finding a
suitable deformation of the action.

Ultimately, we wish to use the supersymmetry to  perform the
integrals, so we must find a deformation of the action which
preserves some supersymmetry. Our aim is to include a mass term for
each field. Our first attempt would be to try to preserve $\delta$
exactly. However, the result of chapter \ref{chap:coho} shows that
then the action must take the form $S = \delta \Tr Q + \epsilon \delta
\Tr R + \mu \Tr W(X_3 + i X_4 )$. One can quickly discover by playing with
$\delta$ that it is not possible to generate mass terms in this
way. Therefore, we must actually deform the supercharge itself.

Lets introduce a deformation parameter $\epsilon$. The
deformed action will be
\beq
\label{eq:modaction}
\Sdef = S + \epsilon S_{1} + \epsilon^{2} S_{2} + \cdots
\eeq
where the $S_{i}$ are all gauge invariant.
The simplest possible modification of $\delta$ is
\beq
\delbar = \delta + \epsilon T
\eeq
where $T$ is some operator.
Then
\beq
\label{eq:moddelsqr}
\delbar^{2} = \delta^{2} + \epsilon \{ \delta , T \} + \epsilon^{2}
T^{2}
\eeq
Our aim is to preserve the supersymmetry, so that 
\beq\delbar \Sdef =
0
\eeq 
Since $\delta^{2} = 0 $ on gauge invariant quantities,
\beq
\delbar \Sdef = 0 \;\;\; \Rightarrow \;\;\; \delbar^{2} \Sdef = 0
\;\;\; \Rightarrow \;\;\; \{ \delta ,T \} S = 0
\eeq
and so $\{ \delta , T \}$ generates one of the continuous bosonic
symmetries of $S$. These are the gauge transformations and $SO(D=4)$
rotations. We follow \cite{Moore:1998et, src:kazakov}\footnote{These
authors apply the localisation method to the cohomological theory
where, although $SO(4)$ is broken,
this $SO(2)$ is still a symmetry.} and use an
$SO(2)$ subgroup of the $SO(4)$
\beq
\begin{array}{ll}
\mathcal{U}: & X_{a} \rightarrow \epsilon_{ab} X_{b} \\
   & \psi_{a} \rightarrow \epsilon_{ab} \psi_{b}
\end{array}
\eeq
to set
\beq\label{T1}
\{ \delta , T \} = \mathcal{U}.
\eeq
Now $\mathcal{U}$ is a compact symmetry which we would like to
preserve, so we impose
\beq
\{ \delta , T \} S_{i} = 0, \;\;\;i=1,\cdots
\eeq
Then (\ref{eq:modaction}) and (\ref{eq:moddelsqr}) imply
\beq
T^{2} S = T^{2} S_{i} = 0, \;\;\; i=1,\cdots
\eeq
and so we shall require
\beq\label{T2}
T^2=0.
\eeq
The simplest possible
form for $T$ is linear
\beq\label{E29}
TA^a = \alpha^{ab}A^b
\eeq
where $A^a$ are the matrix fields and $\alpha^{ab}$ are some
parameters. Imposing \ref{T1} and \ref{T2} on \ref{E29} gives two
possibilities
\begin{itemize}
\item[(i)]
\beq
\begin{array}{ccccccc}
T X_{a} & = & 0 & & T \psi_{a} & = & i \epsilon_{ab} X_{b} \\
T X_3 & = & -i\nu \eta_{1} & & T \eta_{2} &=& - i\nu H \\
T \eta_{1} & = & 0 & & TH & = & 0 \\
T X_4 & = & \nu \eta_1
\end{array}
\eeq
\item[(ii)]
\beq
\begin{array}{lllllll}
T X_{a} & = & 0 & & T \psi_{a} & = & i \epsilon_{ab} X_{b} \\
T X_3 & = & 0 & & T \eta_{2} &=& 0 \\
T \eta_{1} & = & -i \gamma \half (X_3 - i X_4 ) -i \lambda \half (X_3
+ i X_4 ) & & TH & = & i
\gamma \eta_{2} \\
T X_4 & = & 0
\end{array}
\eeq
\end{itemize}
where $\nu$, $\gamma$ and $\lambda$ are complex parameters.

The first possibility is not helpful for generating mass terms in
$X_3$ and $X_4$, so we consider only the second possibility.
Then we have arrived at an altered supercharge $\delbar = \delta +
\epsilon T$
which we hope to be able to preserve as a symmetry of a regularised
action.
\beq\label{delbar}
\begin{array}{l}
\begin{array}{lllllll}
\delbar X_{a} & = & \psi_{a} \, , & & \delbar \psi_{a} & = & \half [X_3 + i
X_4,X_{a}]
+i\epsilon \, \epsilon_{ab} X_{b} \\ 
\delbar \, X_3 & = & \eta_{2} \, , & & \delbar \eta_{2} &=& {i \over 2}[X_4,X_3] \\
\end{array}
\\
\begin{array}{lll}
\delbar \eta_{1} & = & H -i {\epsilon \over 2} \left\{ \gamma (X_3 - i X_4 ) +
\lambda (X_3 + i X_4 ) \right\}  \\
 \delbar
H & = & \half [X_3+iX_4,\eta_{1}] +  i \epsilon
\gamma \eta_{2}\\
\delbar X_4 & = & i \eta_2
\end{array}
\end{array}
\eeq
This modified supercharge $\delbar$ has the property
\beq
\delbar^2 = \delta^2 + \epsilon \, \C U
\eeq
The first row of \ref{delbar} is the part which generates the rotation $\C
U$. This corresponds to the deformation used in \cite{Moore:1998et}
and \cite{src:kazakov} for the cohomological theory. However, we have
also
 added the terms in the third row of \ref{delbar} in order
to generate some useful mass terms for $X_3$ and $X_4$.

On $\C U$-invariant quantities, $\delbar^{2} = \delta^{2}$. This
gives three particularly useful identities:
\begin{itemize}
\item[(i)] $\delbar^{2} =\delta^{2}$ on quantities independent of
$X_{\alpha}, \psi_{\alpha}$
\item[(ii)] $\delbar^{2}(\psi_{\alpha} X_{\alpha}) = \delta^{2}
(\psi_{\alpha} X_{\alpha})$
\item[(iii)] $\delbar^{2}(\epsilon^{\alpha \beta} \psi_{\alpha}
X_{\beta}) = \delta^{2} (\epsilon^{\alpha \beta} \psi_{\alpha}
X_{\beta})$
\end{itemize} 
We are now ready to define the deformed action. Recall that the
original action is $S= \delta \Tr Q$ where $Q$ is defined in equation
\ref{Q}. We define
\beq\label{Sdeforig}
\Sdef = \delbar \Tr Q - i \kappa_1 \delbar \Tr R_1 - i \kappa_2 \delbar
\Tr R_2 -{\mu^2 \over 4} \Tr (X_3 + i X_4)^2.
\eeq
Using the identities above, we note that $Q$ is $\C U$-invariant. If
we also choose $R_1$ and $R_2$ to be $\C U$-invariant then $\Sdef$
will satisfy
\beq
\delbar \Sdef =0
\eeq
so that $\delbar$ is a supersymmetry of the deformed action.
Further, the original action $S$ will be recovered in the limit $\epsilon,
\kappa_1, \kappa_2, \mu \rightarrow 0$.
Specifically, we choose
\beq
R_1 = \half \epsilon_{ab} \psi_{a} X_{b}
\eeq
and
\beq\label{R2}
R_2 = \half \eta_1 (X_3 - i X_4 )
\eeq
which are again both $\C U$-invariant by the above identities.

For reference, the complete deformed action given by this prescription
is given in appendix \ref{app:def}. However, for our present purpose,
we shall find it useful to choose parameters
\beq
\begin{array}{lcr}
\lambda&=&0\\
\gamma&=&-3\\
\kappa_1&=&\epsilon.
\end{array}
\eeq
and we also impose
\beq\label{ineq}
\mu^2 < 3 \epsilon \kappa_2.
\eeq
Then the deformed action is
\bea
\Sdef &=& S \nonumber \\
&&+\epsilon  \Tr \left( X_4\comm{X_1}{X_2}
+3{i \over 2}(X_3-iX_4)H +3i
\eta_1 \eta_2 + i \psi_1 \psi_2  +{\epsilon \over 2} (X_1^2 + X_2^2) \right) \nonumber
\\ &&
+ \kappa_2 \Tr \left( -{i \over 2} H (X_3 -iX_4) +{3\epsilon \over
4} (X_3-iX_4)^2  + i
\eta_1 \eta_2 \right) \nonumber
\\ && \label{defac}
-\mu^2 \Tr \left( \half (X_3 +i X_4) \right)^2
\eea
We now consider the deformed partition function
\beq\label{intorder}
\Zdef = \int dX_4 dX_1 dX_2 dH dX_3 d\psi_1 d\psi_2 d\eta_2 d\eta_1
e^{-\Sdef}.
\eeq
We must first check that this is a convergent integral.
Performing the integrals over the fermions generates a polynomial
which is the deformed version of the Pfaffian, so we write
\beq\label{intorder2}
\Zdef = \int dX_4 dX_1 dX_2 dH dX_3 \, \Pfdef(X_\sigma) \, \exp
\left(-\Sdef|_{\psi=\eta=0} \right).
\eeq
We shall always perform the integrals in the order indicated by the
measures of \ref{intorder} and \ref{intorder2}, so we begin by
considering the integral over $X_3$. We see from \ref{defac} that the
integrand is exponentially damped in $X_3$ as long as the condition
\ref{ineq} holds.
This means we 
can change variables from $X_3$ to
\beq
\phibar = -\half (X_3-iX_4)
\eeq
and then follow the same procedure as before\footnote{That
is, the procedure we used to change variables from $D$ to $H$ and then
take $H$ hermitian.} to  shift the
contours of integration so that $\phibar$ becomes
\begin{it}hermitian\end{it}, and the measure becomes
\beq
dX_3 = 2^{N^2-1} d \phibar.
\eeq
One might worry that after doing some of
the integrations there may be some poles that we should pick
up. However, since the integrand is continuous, we can always change
the order of integration so that the contour we are shifting is the
first, and in that case we do not have to worry about any poles since
the integrand is analytic.

\noindent
In this new picture, the deformed action is
\bea
\Sdef &=& S \nonumber \\
&&+\epsilon  \Tr \left( X_4\comm{X_1}{X_2}
-3i \phibar H +3i
\eta_1 \eta_2 + i \psi_1 \psi_2  +{\epsilon \over 2} (X_1^2 + X_2^2) \right) \nonumber
\\ &&
+ \kappa_2 \Tr \left( i  \phibar H  +3\epsilon  \phibar^2  + i
\eta_1 \eta_2 \right) \nonumber
\\ &&
-\mu^2 \Tr \left( i X_4 - \phibar \right)^2 \label{Sdefphibar}
\eea
where
\bea
S &=& \Tr
\{
H^{2} + H\comm{X_{1}}{X_{2}} +i[X_{a},X_4] [X_{a},\phibar] -
\comm{X_a}{\phibar}^2  - [X_4,
\phibar]^{2} \nonumber  \\ &&  \;\;\;\;\;\;  -\epsilon^{a b} \eta_{1}
[\psi_{a} ,X_{b}]
+\eta_{2} [\psi_{a} ,X_{a} ] - \eta_{a} \comm{iX_4 - \phibar}{ \eta_{a}}
- \psi_{a} 
\comm{\phibar}{ \psi_{a}}  \}
\eea
and the deformed partition function
\beq\label{zdefnew}
\Zdef = \int dX_4dX_1dX_2dHd\phibar \, \Pfdef \, \exp
\left(-\Sdef|_{\psi=\eta=0} \right).
\eeq
It is now easy to see that $\Zdef$  is convergent, since \ref{zdefnew}
is absolutely convergent. To check this, we  examine the real
part of the exponent:
\beq\label{testagain}
\real (\Sdef |_{\psi=\eta=0}) > \Tr \left[ H^2 +{\epsilon^2 \over 2}
(X_1^2 + X_2^2 ) + 3 \epsilon \kappa_2 \phibar^2 + \mu^2 X_4^2 - \mu^2
\phibar^2 \right],
\eeq
where the inequality is obtained by dropping some positive terms from
$S$. 
Then since we assumed $3\epsilon \kappa_2 >\mu^2$, we see that the
deformed partition function \ref{intorder} is a manifestly convergent
integral. 

\section{Integration by Parts}
\label{parts}
In the current scheme, the deformed supercharge (\ref{delbar}) is
\beq\label{delbarphibar}
\begin{array}{lllllll}
\delbar X_{a} & = & \psi_{a} & & \delbar \psi_{a} & = &  [ i
X_4 -\phibar ,X_{a}]
+i\epsilon \, \epsilon_{ab} X_{b} \\ 
\delbar \, \phibar & = & -\eta_{2} & & \delbar \eta_{2} &=&
-i[X_4,\phibar] \\  
\delbar \eta_{1} & = & H -3i \epsilon   \phibar  & & \delbar
H & = & [iX_4 - \phibar,\eta_{1}] -3  i \epsilon
\eta_{2}\\
\delbar X_4 & = & i \eta_2
\end{array}
\eeq
As short hand, lets write $A^a$ for the matrix fields so that $A^1=X_4$,
$A^2=X_1$, and so on. We write $dA$ for the measure $dA= dA^1
\cdots dA^9$. Then the deformed matrix integral \ref{intorder} has
become
\beq
\Zdef (\epsilon, \kappa_2, \mu) = \int dA e^{-\Sdef}.
\eeq
Differentiating with respect to $\kappa_2$ gives
\beq
{\partial \over \partial \kappa_2} \Zdef (\epsilon, \kappa_2, \mu) = -i
\int dA \,
 \delbar (\Tr R_2 ) \, e^{-\Sdef} 
\eeq
where 
\beq
R_2 = -\eta_1 \phibar
\eeq
as we see from the definitions \ref{Sdeforig} and \ref{R2}. This
integral is also manifestly convergent because of the exponential
vanishing of $e^{-\Sdef}$. Since $\delbar \Sdef=0$ we have
\beq
{\partial \over \partial \kappa_2} \Zdef (\epsilon, \kappa_2, \mu) = -i
\int dA \,
 \delbar \left( \Tr R_2  \, e^{-\Sdef} \right) 
\eeq
As before, we write the supercharge \ref{delbarphibar} as a
differential operator
\beq
\delbar = (\delbar A^a ) {\partial \over \partial A^a}.
\eeq
We note from \ref{delbarphibar} that each $\delbar A^a$ is independent
of the matrix field $A^a$. Thus we can write
\beq
\delbar \; \cdot = (\delbar A^a ) {\partial \; \cdot \over \partial A^a} 
= {\partial \over \partial A^a} \left( \delbar A^a \, \cdot \right)
\eeq
and so
\beq
{\partial \over \partial \kappa_2} \Zdef (\epsilon, \kappa_2, \mu) = -i
\int dA \,
{\partial \over \partial A^a}  \left( \delbar A^a \Tr R_2  \,
e^{-\Sdef} \right).  
\eeq
We consider each term in the sum over $a$ in this expression
separately. 

If $A^a$ is fermionic, then each $A^a_{ij}$ is
Grassman. Differentiating $\partial / \partial A^a_{ij}$ removes
$A^a_{ij}$ from the integrand, and so the integral is identically zero.

If $A^a$ is bosonic, then since the integrand vanishes exponentially,
we also get zero by the divergence theorem. So we have found
\beq
{\partial \over \partial \kappa_2} \Zdef (\epsilon, \kappa_2, \mu) = 0;
\eeq
the deformed partition function is independent of $\kappa_2$. Our
tactic will be to evaluate the integral when $\kappa_2$ is large.

The part of the deformed action (\ref{Sdefphibar}) which depends on
$\kappa_2$ is
\beq\label{eq:2.58}
\kappa_2 \Tr ( i  \phibar H  +3\epsilon  \phibar^2  + i
\eta_1 \eta_2 ) = \kappa_2 \Tr ( 3\epsilon  (\phibar+
{i \over 6 \epsilon} H)^2 + {1 \over 12 \epsilon} H^2 + i
\eta_1 \eta_2 ).
\eeq
We shall take $\kappa_2$ large, and use a saddle point method to
integrate out $\phibar$. To avoid breaking the flow of argument here,
a full description of the relevant saddle point method is included in
appendix \ref{AppC}. The first step is to follow the by now
familiar technique, and set $\phibar' = \phibar +
{i \over 6 \epsilon} H$, and shift the contours so that
$\phibar'$ becomes hermitian. In order to apply the saddle
point method, we use equations \ref{tilde1}-\ref{tilde2} to change
to tilde type variables, and then apply the method to each of these
real integration variables. Then we find $\phibar'$ localises at $-{i
\over 6 \epsilon} H$, and $H$ localises at $0$. Thus
\beq
\int d \phibar d H g( \phibar , H) \exp \left\{ - \kappa_2 \Tr ( i  \phibar H
+3\epsilon  \phibar^2   ) \right\} = {2^{N^2-1} \over N} {\pi^{N^2-1} \over
(\kappa_2)^{N^2-1}} \, g(0,0) + 
\C O \left( \kappa_2^{-N^2} \right)
\eeq
where the $2^{N^2-1}$ factor comes from  the factors $3\epsilon$ and
$(12 \epsilon)^{-1}$ which appear in equation \ref{eq:2.58}.

We would now like to integrate out $\eta_1$ and $\eta_2$.
Consider
\beq
I(\kappa_{2}) = \int d\eta_2 d \eta_1 f(\eta_1 ,\eta_2 ) \exp
(-i\kappa_{2} \Tr 
\eta_1 \eta_2)
\eeq
for some $f$. Changing variables to $\widetilde{\eta_1}
,\widetilde{\eta_2}$, this is
\beq
I(\kappa_{2}) = \int d \widetilde{\eta_2} d \widetilde{\eta_1}
f(\widetilde{\eta_1} , \widetilde{\eta_2}) \exp 
(-i\kappa_{2} \Tr \widetilde{\eta_1}^{T} \widetilde{\eta_2} )
\eeq
and using $(\Tr \widetilde{\eta_1}^{T} \widetilde{\eta_2} )^{N^{2}}=0$
(since $\widetilde{\eta_1}$ and $\widetilde{\eta_2}$ are traceless
real $\NbyN$ Grassman matrices),
\beq\label{eq:2.62}
I(\kappa_{2})  =  i^{N^{2}-1} \int d \widetilde{\eta_1} d
\widetilde{\eta_2} f(0,0)
{\kappa_{2}^{N^{2}-1} \over (N^{2}-1)!} (\Tr \widetilde{\eta_1}^{T} 
\widetilde{\eta_2} )^{N^{2}-1} + \mathcal{O}(\kappa_{2}^{N^{2}-2}). 
\eeq
Note that we also switched the order of integration of $\eta_1$ and
$\eta_2$ in order to cancel the $(-1)^{N^2-1}$ which came from
expanding the exponential.
At present, we are integrating over traceless $\widetilde{\eta_1}
,\widetilde{\eta_2}$. It is simpler to do the integration if we insert
\beq
\deltafn{(\Tr \widetilde{\eta}_1 )} =  (\widetilde{\eta}_1)_{11} +
\cdots + (\widetilde{\eta}_1)_{NN}
\eeq
and similar for $\eta_2$, and integrate over the full matrices. If we
do this, and also expand the trace in \ref{eq:2.62}, we find
\bea
I(\kappa_{2})  &=&  i^{N^{2}-1} \int d \widetilde{\eta_1} d
\widetilde{\eta_2} f(0,0)
\kappa_{2}^{N^{2}-1}  N \prod_{i,j} \widetilde{\eta_1}_{ij}  
\widetilde{\eta_2}_{ij} + \mathcal{O}(\kappa_{2}^{N^{2}-2}) \\
&=& \kappa_{2}^{N^{2}-1} i^{N^{2}-1} N f(0,0) +
\mathcal{O}(\kappa_{2}^{N^{2}-2}) 
\eea
We can now use these results to integrate out $\phibar$, $H$, $\eta_1$
and $\eta_2$ from the deformed partition function
\bea
\Zdef &=& \int dX_4 dX_1 dX_2 d \psi_1 d\psi_2 d\eta_1 d\eta_2 dH
(2^{N^2-1}d\phibar ) \exp(-\Sdef) \\
&=& (4i\pi)^{N^2-1} \int dX_4 dX_1 dX_2 d \psi_1 d\psi_2 \exp (-\Sdef
)|_{\eta_1=\eta_2=0, \phibar = H=0} + \C O ( \kappa_2^{-1} ).\nn
\eea
However, since we know $\Zdef$ is independent of $\kappa_2$, the $\C O
( \kappa_2^{-1} )$ terms must actually be zero. Writing this out
in full gives
\bea
\Zdef &=& (4i\pi)^{N^2-1} \int dX_4 dX_1 dX_2 d \psi_1 d\psi_2
\nonumber \\ && \exp \Tr
\left(- \epsilon X_4 \comm{X_1}{X_2} - \epsilon i \psi_1 \psi_2 -
{\epsilon^2 \over 2} (X_1^2 + X_2^2) - \mu^2 X_4^2 \right)
\eea
and we note here that we can scale out a factor of $\epsilon$ from the
first three terms in the exponential. This process of integrating out
$\phibar, H, \eta_1, \eta_2$ in the large $\kappa_2$ limit has
sometimes been
known as integrating over a BRST quartet.

The fermions $\psi_1$, $\psi_2$ can be integrated out immediately
using \ref{typferm} (but note that the sign difference gives an extra
factor $(-1)^{N^2-1}$). We can also integrate out $X_2$ by completing
the square and using \ref{tilde2}, to give
\beq\label{3:2.69}
\Zdef = \sqrt{N} \left( { 2 \pi \over \epsilon } \right)^{N^2 -1
\over 2} (4 \pi )^{N^2 -1} \int d X_4 d X_1 \exp \Tr \left( {1 \over 2
\epsilon } \comm{X_4}{X_1}^2 -{\epsilon \over 2} X_1^2 - \mu^2 X_4^2
\right).
\eeq
We now follow the usual procedure to reduce the integral over $X_4$ to
an integral over its eigenvalues (see for example
\cite{src:randmatrices} for an 
account of this method applied to the $SU(N)$ groups). For a function
$f$ which depends only on the eigen values $x_1, \cdots , x_N$ of
$X_4$, we have
\beq
\int dX_4 f(x_1, \cdots , x_N ) = c_N \int dx_1 \cdots d x_N
\prod_{j<k} (x_j - x_k )^2 f(x_1 , \cdots , x_N ).
\eeq
Recall that for $X_4$ only, we are using the normalisation $ \int dX_4
\deltafn{(\Tr X_4 )} \, e^{-2\Tr X_4^2}=1$ (equation \ref{X4norm}). In
this case, the constant is 
\beq
c_N =    {2^{N^2} \, \sqrt{N \pi /2} \over  (2 \pi )^{N \over 2} \, 1! 2!
\cdots N!}
\eeq
as can most easily be seen by adjusting the value given in
\cite{src:randmatrices} to our conventions.
It is now convenient to use the tilde notation of
equations \ref{tilde1}-\ref{tilde2}, $X_1 \rightarrow \widetilde{X}$,
since then
\beq
\Tr \comm{X_1}{X_4}^2 = - \sum_{i,j} (\widetilde{X}_{ij})^2 (x_i
-x_j)^2.
\eeq
Then \ref{3:2.69} becomes
\bea
\Zdef &=& \sqrt{N} \left( {2 \pi \over \epsilon } \right)^{N^2 -1 \over
2} (4\pi)^{N^2-1} c_N \int dx_1 \cdots dx_N \prod_{j<k} (x_j -x_k)^2
d\widetilde{X} \deltafn{(x_1 + \cdots + x_N)} \nonumber \\ && \exp
\left( -{1 \over 2 
\epsilon} \sum_{i,j} (\widetilde{X}_{ij})^2 \left\{ (x_i-x_j)^2 +
\epsilon^2 \right\} - \mu^2 \sum_i x_i^2 \right).
\eea
It is convenient to define
\beq
\C F_N = { 2^{N(N+1) \over 2} \pi^{N-1 \over 2} \over 2 \sqrt{N}
\prod_{i=1}^{N-1} i!}
\eeq
since this is the normalisation constant that emerged in the numerical
calculations of the partition function of \cite{Krauth:1998xh,
Krauth:1998yu}. Then, integrating out $\widetilde{X}$, we are 
left with
\bea\label{result1}
\Zdef &=& (2 \pi)^{2(N^2-1)} \, 2^{{3 \over 2}(N^2-1)} \, {\C F_N \over (2\pi
\epsilon)^{N-1} (N-1)!} \\&&  \times \int dx_1 \cdots dx_N \, \deltafn{(x_1 +
\cdots + x_N)} \prod_{i > j} { (x_i-x_j)^2 \over (x_i - x_j)^2 + \epsilon^2}
\, e^{-\mu^2 \sum_i x_i^2}. \nonumber
\eea
This result is almost identical to the formula of MNS up to 
normalisation. To get back to
their formula, we would send $\mu \rightarrow 0$, and change the sign of the
$\epsilon^2$ which appears in the denominator of the product (this
latter difference is an advantage since MNS added an imaginary
part to $\epsilon$ by hand as part of their prescription for
performing the contour integrals).
As it stands, the formula \ref{result1} diverges in the limit
$\epsilon \rightarrow 0$ or $\mu 
\rightarrow 0$.

\section{Dangerous Fermion Masses}
\label{fermmass}
Lets summarise. We began by considering the
supersymmetric Yang-Mills integral \ref{eq:zzz}. We added small mass
terms (and a small cubic term) to the action leading to the deformed
action \ref{defac}. After checking that the resulting deformed
partition 
function $\Zdef$ is convergent, we observed that it is independent of
one of the small parameters $\kappa_2$. Then we followed the 
localisation method, and calculated the integral by taking $\kappa_2$
large. 

We hoped that we could then take the small parameters to zero, and
arrive back at the original Partition function. However, this is
clearly not the case since if we do this in equation \ref{result1}, we
get $\infty$, whilst we have shown in chapter \ref{chap:conv} that the
result should be finite.

A rough calculation  indicates why this may be so. We have
attempted to 
deform the action by adding mass terms for the bosons and fermions
\beq\label{model}
\Sdef \sim S +i \epsilon \Tr \left( \psi_1 \psi_2 + \eta_1 \eta_2
\right) + \epsilon^2 \Tr \left( X_\mu X_\mu \right).
\eeq
We did not quite achieve this in equation \ref{defac} since the mass
term for $X_4$ has the wrong sign. However, rewriting \ref{defac} in
the form \ref{Sdefphibar} shows that \ref{model} is a fair model. Lets
go back to the convergence argument of section \ref{superconv}. The
new fermion mass terms introduce extra insertions into the modified
pfaffian \ref{modpfaff}. However, these insertions do
\begin{it}not\end{it} come with associated powers of $R^{-(2-\eta)}$
as they do in the convergence proof,
but they do each have a power of $\epsilon$. 
In particular, consider the term in which all the $2(D-2)$ $J$-type
insertions are present (the $\xi_\alpha$). Then we loose a factor
$R^{-2(2-\eta)(D-2)} \sim R^{-4(D-2)}$ from the convergence proof, and
also an 
$R^{D-2}$ from the measure, but gain $\epsilon^{D-2}$. So, without the
$\epsilon^2 R^2 \Tr x_\mu x_\mu$ mass terms, we would have
\beq\label{tester}
\int {dR \over R}R^{Dg} R^{(D-2)g} R^{-(D-2)} \C I_{D,G} \sim \epsilon^{D-2}
\int {dR \over R} R^D .
\eeq 
We have not yet considered the bosonic mass terms which appear in the
exponential
\beq
\exp( -\epsilon^2 R^2
\Tr x_\mu x_\mu).
\eeq
We would like to find the best possible bound on the
$\epsilon$-behaviour of the partition function. Thus, we would like to
use the bosonic mass terms \begin{it}just\end{it} enough to make
\ref{tester} convergent. We can bound
\beq
\exp( -\epsilon^2 R^2
\Tr x_\mu x_\mu) < \exp( -\epsilon^2 R^2
\Tr \rho_\mu \rho_\mu)
\eeq
and then, for fixed $\epsilon$ and large $R$, the integrations over
the $\rho_\mu$ in the convergence proof lead to an additional
$(R\epsilon)^{-D}$. If we add this into the rhs of \ref{tester}, we see
that this sets the $R$-integration on the threshold of
convergence. Thus, the best bound we can get on $\Zdef$ using the
methods we have developed is
\beq
\Zdef \sim \epsilon^{-2}.
\eeq
Thus, the convergence proof (which we found gave perfect predictions for
convergence of the undeformed partition function) fails for the
deformed partition function when $\epsilon \rightarrow 0$.
Of course, this does not prove anything since this only gives a crude
\begin{it}upper\end{it} bound on certain terms of $\Zdef$. However, it
does indicate
very clearly how after adding fermion mass terms to an action, we
cannot expect to regain the partition function by sending the masses
back to zero. In the case of $SU(2)$ where no iteration is involved,
there is a decent chance for this basic argument to give the correct power
of $\epsilon$. Indeed, setting $\mu=\epsilon$ in \ref{result1}, and
sending $\epsilon \rightarrow 0$, we obtain $\Zdef \sim \epsilon^{-2(N-1)}$
for $SU(N)$, and so $\epsilon^{-2}$ for $SU(2)$.

Since the limit of masses going to zero is dangerous, one might also
worry about taking the limit $\kappa_2 \rightarrow \infty$ in the
arguments of the previous section. We know that the integral is
independent of $\kappa_2$ after doing the integration, but we don't
know whether we can take $\kappa_2 \rightarrow \infty$ before
integrating.  However in reality, we do not take
the limit as $\kappa_2 \rightarrow \infty$ first, but rather find the first
term in an asymptotic expansion, and use the result of appendix
\ref{AppC} to show that the correction term is finite, and vanishes as
$\kappa_2 \rightarrow \infty$ and so is zero. 

\section{An Unresolved Issue}
We recall that in \cite{Moore:1998et}, the deformation method was
applied to the cohomological theory. The result  is
almost identical to the equation \ref{result1} that we found by
applying the method to the Yang-Mills theory. In \cite{Moore:1998et},
this result is found as an integral over the eigenvalues of the
cohomological theory field $\phi$, whilst we have obtained it as an
integral over the eigenvalues of $X_4$. The
authors of \cite{Moore:1998et} integrated out the $\delta$-function,
and then set a prescription for completing the contours of the
$N-1$ remaining integrals  around either the upper or lower half
plane. They were able to perform the contour integrals, and found
\beq
{1 \over  (2\pi
\epsilon)^{N-1} (N-1)!} \oint dx_1 \cdots dx_{N-1} \int dx_N \,
\deltafn{(x_1 + 
\cdots + x_N)} \prod_{i \neq j} { x_i-x_j \over x_i - x_j + i\epsilon}
= {1 \over N^2}.
\eeq
They also applied the same method to $D=6$ and $D=10$ theories, and
noticed that for $D=10$ the result corresponds to the conjecture of
Green and Gutperle
\cite{Green:1998tn}. At the same time, Krauth, Nicolai and Staudacher
\cite{Krauth:1998xh} 
were able to apply Monte Carlo techniques to evaluate the partition
function for some values of $N$ and obtained the same result for $D=4, 6,
10$, and also calculated the normalisation factor $\C F_N$ as a group
volume. 

It is interesting to understand exactly why the group volume
$\C F_N$ appears as the normalisation factor for the MNS
formula. Therefore, it is worthwhile to adapt our conventions to those
that were used for the numerical calculations
\cite{Krauth:1998xh,Krauth:1998yu}, and check agreement. For $D=4$,
these authors calculate 
the 
partition function defined as
\beq
\C Z'_{4,N}=\int \prod_{a=1}^{N^2-1} \left(\prod_{\mu=1}^4 {dX_\mu^a \over
\sqrt{2\pi}} \right) \left( \prod_{\alpha=1}^4 d \Psi_\alpha^a \right)
\exp \left[ \half \sum_{\mu,\nu} \Tr \comm{X_\mu}{X_\nu}^2 + \Tr
\Psi_\alpha \comm{\Gamma^\mu_{\alpha \beta} X_\mu}{\Psi_\beta}
\right]
\eeq
where the fermions are written in the real representation of equation
\ref{intro:partn}. 
Comparing to \ref{eq:zzz}, this differs from our definition by a factor
$(2\pi)^{-2(N^2-1)}$ in the 
measure, and a factor $\half$ from our potential $\quarter
\comm{X_\mu}{X_\nu}\comm{X_\mu}{X_\nu}$. We can get back to our action
\ref{ymaction2} by scaling $X_\mu
\rightarrow 2^{-\quarter}X_\mu$ and $\Psi_\alpha \rightarrow 2^{1
\over 8} \Psi_\alpha$.
Then we pick up an additional factor $(2^{-\quarter})^{4(N^2-1)} \,
(2^{1 \over 8})^{-4(N^2-1)} = 2^{-{3 \over 2}(N^2-1)}$ in the
measure. Then adding these factors into the result \ref{result1}, and
sending $\mu \rightarrow 0$,
leads to
\beq
 {\C F_N \over (2\pi
\epsilon)^{N-1} (N-1)!}  \int dx_1 \cdots dx_N \, \deltafn{(x_1 +
\cdots + x_N)} \prod_{i > j} { (x_i-x_j)^2 \over (x_i - x_j)^2 +
\epsilon^2}  \nonumber
\eeq
so that applying the MNS contour prescription gives
\beq
\C Z'_{4,N}=\C F_N {1 \over N^2}
\eeq
in exact agreement with the numerical calculations.

It was hoped that once the relation between the cohomological
calculation of \cite{Moore:1998et} and the Yang-Mills model was
understood, the reason for requiring the contour integrals would
become readily  
apparent. Sadly, although through the work of this chapter we can now
understand the calculation 
entirely from the Yang-Mills perspective, this has not come to
pass. It is clear from section \ref{fermmass} that the problem with
the calculation as it stands is in the deformation of the
action. Therefore, the task ahead is to find a new deformation that
will still allow all the various steps of the previous calculation,
but not suffer the problems of the current deformation. For the
present, we must leave this as an unresolved issue.

\chapter*{Summary of Results} 
\addcontentsline{toc}{chapter}  
		 {\protect\numberline{Summary of Results\hspace{-196pt}}} 
In chapter \ref{chap:conv} we considered the convergence properties of
Yang-Mills matrix models. For the bosonic theories, we  showed that
the partition function 
converges when $D \geq D_c$ and calculated $D_c$ for each  group. We
also calculated a critical value $k_c$ for each group with the
property that any correlation function of degree $k<k_c$ is
convergent. Conversely, we showed that there is always a correlation
function of degree 
$k_c$ which is divergent.

For the supersymmetric models, we showed that the partition function
converges when $D=4,6$ and $10$, and that correlation functions of
degree $k<2(D-3)$ are convergent. This result applies for any
compact semi-simple gauge group.

In chapter \ref{chap:coho} we considered the supersymmetric
models. With particular reference to the $D=4$ model, we found all
quantities invariant under the supercharge. We also indicated how the
result extends immediately to $D=6$ and $D=10$. In appendix
\ref{app:super}, we point out that all four supercharges in the $D=4$
model are related by permutations of the fields. Thus the result can
immediately be applied to any of the supercharges.

In chapter \ref{chap:3} we considered how to apply the deformation
method of \cite{Moore:1998et} directly to the supersymmetric
Yang-Mills  model with $D=4$.  We found
a deformation of the action that can 
generate mass 
terms for all the fields and still preserve some supersymmetry. This
allowed us to integrate over a BRST quartet rigorously, and confirm the
formula of \cite{Moore:1998et}. We showed why this
method fails to reproduce exact values for the partition function so
that an alternative regularisation must be 
found. However, a proof that the contour prescription of Moore,
Nekrasov and Shatashvili is the correct
regularisation remains elusive.

\renewcommand{\chaptername}{Appendix}
\appendix
\chapter{Supercharges for $D=4$}
\label{app:super}
In this appendix we give an explicit representation of the four
linearly independent supercharges of 
the $D=4$ supersymmetric Yang-Mills matrix theory. They are obtained
directly from the supersymmetry \ref{SS}. We use the notation of
chapter \ref{chap:coho}, and in addition, define $-H'= H+
\comm{X_1}{X_2}$. Although they are written in terms of fields $\phi$ and
$\phibar$ from the Cohomological theory, one can simply make the
replacement \ref{phidef} to return explicitly to the Yang-Mills theory.

\[
\begin{array}{lclclcl}
\delta_{1} X_{a}& =&  \psi_{a} &&
\delta_{1} \psi_{a} & =&  [\phi,X_{a}] \\
\delta_{1} \phibar & =&  -\eta_{2} && 
\delta_{1} \eta_{2} & =&  -[\phi ,\phibar ] \\
\delta_{1} \eta_{1} & =&  H & &
\delta_{1} H & =&  [\phi , \eta_{1} ] \\
\delta_{1} \phi & =&  0 
\\
\\
\delta_{2} X_{a}& =&  -\epsilon_{a b} \psi_{b} &&
\delta_{2} \psi_{a} & =&  \epsilon_{a b} [\phi,X_{b}] \\
\delta_{2} \phibar  &=&  \eta_{1} && 
\delta_{2} \eta_{1}  &=&  [\phi ,\phibar ] \\
\delta_{2} \eta_{2}  &=&  H && 
\delta_{2} H  &=&  [\phi , \eta_{2} ] \\
\delta_{2} \phi  &=&  0 
\\
\\
\delta_{3} X_{a} &=&  \eta_{a} &&
\delta_{3} \eta_{a}  &=&  [\phibar,X_{a}] \\
\delta_{3} \phi  &=&  -\psi_{2} && 
\delta_{3} \psi_{2}  &=&  -[\phibar ,\phi ] \\
\delta_{3} \psi_{1}  &=&  H^{\prime} && 
\delta_{3} H^{\prime}  &=&  [\phibar , \psi_{1} ] \\
\delta_{3} \phibar & =&  0 
\\
\\
\delta_{4} X_{a} &=&  -\epsilon_{a b} \eta_{b} &&
\delta_{4} \eta_{a}  &=&  \epsilon_{a b} [\phibar,X_{b}] \\
\delta_{4} \phi  &=&  \psi_{1} && 
\delta_{4} \psi_{1}  &=&  [\phibar ,\phi ] \\
\delta_{4} \psi_{2}  &=&  H^{\prime} && 
\delta_{4} H^{\prime}  &=&  [\phibar , \psi_{2} ] \\
\delta_{4} \phibar  &=&  0 
\end{array}
\]
These obey the algebra
\beq
\begin{array}{ll}
\delta_{1}^{2}=[\phi,\; \;] & \delta_{2}^{2}=[\phi,\; \;] \\
\delta_{3}^{2}=[\phibar,\; \;] & \delta_{4}^{2}=[\phibar,\; \;] \\
\{ \delta_{1},\delta_{2} \}=0 & \{\delta_{1},\delta_{3}\} =-[X_{1},\;\;] \\
\{\delta_{1},\delta_{4}\}=[X_{2},\;\;] &
\{\delta_{2},\delta_{3}\}=-[X_{2},\;\;] \\
\{\delta_{2},\delta_{4}\}=-[X_{1},\;\;] & \{\delta_{3},\delta_{4}\}=0
\end{array}
\eeq
and so
\beq
\{\delta_{i},\delta_{j}\}=0
\eeq
on gauge invariant quantities.

As we have pointed out, none of the supercharges in the Yang-Mills
theory can preserve hermiticity.  A nice feature of the
Cohomological theory, in which $\phi$ and $\phibar$ become independent
and respectively antihermitian and hermitian, is that then $\delta_1$
and $\delta_2$ become truly real. However, it is interesting that,
unlike the Minkowski theory, only two of the supercharges preserve
hermiticity in this way in the Cohomological theory.

The action is invariant under certain permutations of the matrix
fields. Defining
\beq
\begin{array}{cc}
\Pi: & \psi_{a} \leftrightarrow \eta_{a} \\
         & H \leftrightarrow H^{\prime}      \\
         & \phi \leftrightarrow \phibar      
\end{array}
\eeq
\beq
\begin{array}{cc}\label{Omegaperm}
\Omega: & \psi_{a} \rightarrow \epsilon_{ab} \psi_{b} \\
         & \eta_{a} \rightarrow -\epsilon_{ab} \eta_{b}     
\end{array}
\eeq
\beq
\begin{array}{cc}
\Sigma: & X_{a} \rightarrow -\epsilon_{ab} X_{b} \\
         & \eta_{a} \rightarrow -\epsilon_{ab} \eta_{b}     
\end{array}
\eeq
one finds $\Pi S = \Omega S = \Sigma S = S$.

The supersymmetries are related to each other by the
permutation symmetries. For example
\beq\label{Omega}
\Omega^{-1} \delta_{1} \Omega = \delta_{2}
\eeq
\beq
\Sigma^{-1} \delta_{1} \Sigma = \delta_{2}
\eeq
\beq
\Pi^{-1} \delta_{1} \Pi = \delta_{3}
\eeq
\beq
\Pi^{-1} \delta_{2} \Pi = \delta_{4}
\eeq
Each supersymmetry can be related to $\delta_{1}$ by a permutation
symmetry. Rather than a model with four supersymmetries, one could
think 
of the theory as a model with one supersymmetry together with some
permutation symmetries.

For interest, it is possible to represent the action \ref{eq:Scoh} in
terms of all four supercharges. There are three ways of doing this:
\beq
\begin{array}{ccr}
S & = & \half \delta_{1} \delta_{2} \delta_{3} \delta_{4} \Tr
X_{1}^{2} \\
& = & \half \delta_{1} \delta_{2} \delta_{3} \delta_{4} \Tr
X_{2}^{2} \\
& = & - \delta_{1} \delta_{2} \delta_{3} \delta_{4} \Tr \phi \phibar
\end{array}
\eeq
Since the $\delta_{i}$ anticommute, these representations render $S$
manifestly invariant under all four supercharges.

\chapter{Deformed Action}
\label{app:def}
The deformed action \ref{Sdeforig} written out in full is given by
\bea\label{fulldef}
\Sdef &=& S \nonumber \\
&&+\epsilon  \Tr \left( {i \over 2}(\gamma+2)(-X_3+iX_4)\comm{X_1}{X_2}
-{i \over 2}\gamma (X_3-iX_4)H \right. \nonumber \\
&& \;\;\;\;\;\; \left. - {i \over 2}\lambda (X_3+i X_4)(H+ \comm{X_1}{X_2}) -i\gamma
\eta_1 \eta_2 \right) \nonumber \\
&&
+ \kappa_1 \Tr \left( i \psi_1 \psi_2 -{i \over 2} (X_3+iX_4)
\comm{X_1}{X_2} +{\epsilon \over 2} (X_1^2 + X_2^2) \right) \nonumber
\\ &&
+ \kappa_2 \Tr \left( -{i \over 2} H (X_3 -iX_4) -{\epsilon \gamma \over
4} (X_3-iX_4)^2 -{ \epsilon \lambda \over 4} (X_3^2 + X_4^2) + i
\eta_1 \eta_2 \right) \nonumber
\\ &&
-\mu^2 \Tr \left( \half (X_3 +i X_4) \right)^2
\eea
where $S$ is the original action
\bea
S &=& \Tr \left( (H + \half \comm{X_1}{X_2})^2 - {1 \over
4} \sum_{\mu>\nu}\comm{X_\mu}{X_\nu}^2 \right.  \\
&&
\left.  - \epsilon_{ab} \eta_1 \comm{\psi_a}{X_b}
-\eta_a \half \comm{ (X_3+iX_4)}{\eta_a} -\psi_a \half \comm{
(-X_3+iX_4)}{\psi_a} + \eta_2 \comm{\psi_a}{X_a} \right). \nonumber
\eea
It is interesting to know
how much supersymmetry is preserved by this deformation. 
Certainly, $\Sdef$ is invariant under the deformed supercharge
$\delbar$ given in equation \ref{delbar}. Since, $\delbar$ was deformed
from the original supercharge $\delta_1$ of appendix \ref{app:super},
lets write $\delbar \equiv \delbar_1$.
We note that
$\Sdef$ is also invariant under the permutation symmetry
$\Omega$ defined in equation
\ref{Omegaperm}. Then, following equation \ref{Omega}, we can define
\beq
\delbar_2 = \Omega^{-1} \delbar_1 \Omega
\eeq
so that
\beq
\delbar_2 \, \Sdef =0.
\eeq
Since $\Sdef$ is not invariant under the permutation $\Pi$, it is not
possible to define deformed versions of $\delta_3$ and $\delta_4$ in
this way. Thus $\Sdef$ is invariant under only two
supercharges compared 
to the four of the original action $S$.
\chapter{Saddle Point Method}
\label{AppC}
In this appendix, we prove the precise form of the saddle point method
that we use in chapter \ref{chap:3}. The proof follows the usual
argument of integration by parts; see for example \cite{src:asymp}.

\noindent
Let $g(t)$ be continuous, twice differentiable and
\beq\label{AppC1}
\int_{0}^{\infty} {dt} \absval{g^{(n)}(t) }
\eeq
converge for $n=0,1,2$.
Then
\beq
\int_{0}^{\infty} {dt} g(t) \exp -kt^{2} = g(0) {\sqrt{\pi} \over 2 \sqrt{k} }
 + R(k) 
\eeq
where
\beq
\absval{R(k)} < {\const \over k}.
\eeq
To see this, we use
integration by parts,
\bea
I(k) & = & \int_{0}^{\infty} {dt} g(t) \exp -kt^{2} \\
     & = & \left[- g(t) \int_{t}^{\infty} \exp (-ku^{2}) du
\right]_{0}^{\infty} + \int_{0}^{\infty}{dt} g^{\prime}(t)
\int_{t}^{\infty} {du} \exp (-ku^{2}) \\
     & = & g(0) \int_{0}^{\infty}{du} \exp (-ku^{2}) +
\int_{0}^{\infty}{dt} g^{\prime}(t) \int_{1}^{\infty}{dv} t \exp (-k
t^{2}v^{2})\\
&=&g(0) {\sqrt{\pi} \over 2 \sqrt{k} }
 + R(k)
\eea
where
\beq
R(k)=\int_{0}^{\infty}{dt} g^{\prime}(t) \int_{1}^{\infty}{dv} t \exp (-k
t^{2}v^{2}).
\eeq
Integrating by parts again,
\bea
R(k) & = & \int_{1}^{\infty} {dv} \left( \left[ g^{\prime}(t) {-1 \over
2kv^{2}}\exp(-kt^{2}v^{2}) \right]_{0}^{\infty} - \int_{0}^{\infty} {dt}
g^{\prime \prime}(t) {-1 \over 2 kv^{2}}\exp(-kt^{2}v^{2}) \right) \nonumber\\
 & = & \int_{1}^{\infty} {dv} {g^{\prime}(0) \over 2kv^{2} } +
\int_{1}^{\infty}{dv}\int_{0}^{\infty}{dt} g^{\prime \prime}(t)
{ \exp(-kt^{2}v^{2}) \over 2kv^{2}} 
\eea
so that
\beq
|R(k)|  \leq  {1 \over k} \int_{1}^{\infty} {dv}
{\absval{{g^{\prime}(0)}} \over 2v^{2}} + {1 \over k} \left(
\int_{1}^{\infty}{dv}{1 \over 2v^{2}} \right) \int_{0}^{\infty}{dt}
\absval{g^{\prime \prime}(t)}  = { \const \over k}
\eeq
There is an issue we must address specifically in relation
to the saddle point calculation of chapter \ref{chap:3}. In this case
we must repeat the saddle point process many times as we integrate out
each variable in the matrices $\phibar$ and $H$. Properly, we should
write $\kappa_2= k+1$. Then setting $k=0$ (i.e. $\kappa_2=1$) gives us
the integrand which corresponds to $g(t)$. This integrand is
exponentially damped, and so the properties \ref{AppC1} always
hold. In particular the integrals which define the remainder term
always converge.

\newpage
 
\singlespacing
\addcontentsline{toc}{chapter} 
                 {\protect\numberline{Bibliography\hspace{-96pt}}} 

\providecommand{\href}[2]{#2}\begingroup\raggedright\endgroup

\end{document}